\input cp-aa.tex
% Paper in A\&A style  Plain TeX source code
%
%
% some definitions for the references
%
%
%
\def\vp #1 #2{ #1, #2}
\def\a #1:#2 #3 #4 {#1, ~#2, {A\&A} \vp #3 #4}
\def\apj #1:#2 #3 #4 {#1, ~#2, {ApJ} \vp #3 #4}
\def\asupl #1:#2 #3 #4 {#1, ~#2, {A\&AS} \vp #3 #4}
\def\apjsupl #1:#2 #3 #4 {#1, ~#2, {ApJS} \vp #3 #4}
\def\pasp #1:#2 #3 #4 {#1, ~#2, { PASP} \vp #3 #4}
\def\paspc #1:#2 #3 #4 {#1, ~#2, { PASPC } \vp #3 #4}
\def\mn #1:#2 #3 #4 {#1, ~#2, {MNRAS} \vp #3 #4}
\def\msait #1:#2 #3 #4 {#1, ~#2, {Mem. Soc. Astron. It.} \vp #3 #4}
\def\nat #1:#2 #3 #4 {#1, ~#2, {Nat} \vp #3 #4}
\def\aj #1:#2 #3 #4 {#1, ~#2, {AJ} \vp #3 #4}
\def\jaa #1:#2 #3 #4 {#1, ~#2, {JA\& A} \vp #3 #4}
\def\aspsc #1:#2 #3 #4 {#1, ~#2, {Ap\&SS} \vp #3 #4}
\def\anrev #1:#2 #3 #4 {#1, ~#2, {ARA\&A} \vp #3 #4}
\def\esomsg #1:#2 #3 #4 {#1, ~#2, {The Messenger} \vp #3 #4}
\def\rmp #1:#2 #3 #4 {#1, ~#2, {Rev. Mod. Phys.} \vp #3 #4}
\def\ans #1:#2 #3 #4 {#1, ~#2, {Ann. Rev. of Nucl. Sci.} \vp #3 #4}
\def\phrev #1:#2 #3 #4 {#1, ~#2, {Phys. Rev.} \vp #3 #4}
\def\phreva #1:#2 #3 #4 {#1, ~#2, {Phys. Rev. A} \vp #3 #4}
\def\phs #1:#2 #3 #4 {#1, ~#2, {Physica Scripta} \vp #3 #4}
\def\jqsrt #1:#2 #3 #4 {#1, ~#2, {J. Quant. Spectrosc. Radiat.
       Transfer} \vp #3 #4}
\def\cjp #1:#2 #3 #4 {#1, ~#2, {Can. J. Phys. } \vp #3 #4}
\def\jphb #1:#2 #3 #4 {#1, ~#2, {J. Phys. B} \vp #3 #4}
\def\apop #1:#2 #3 #4 {#1, ~#2, {Appl. Opt.} \vp #3 #4}
\def\gca #1:#2 #3 #4 {#1, ~#2, {Geochim. Cosmochim. Acta}\vp #3 #4}
%
%
% table  numbering
%
\newcount\tabnumber
\def\cleartabnumber{\global\tabnumber=0}
\def\tbl #1{\global\advance\tabnumber by 1
\tabcap{\the\tabnumber}{#1}}

%
% useful for tables
%

%
% some shorthand
%
\def\kms{$\rm km s^{-1}$}
\def\eby{$\rm E(b-y)$}

\def\t{${T}_{\rm eff}~$}
\def\g{$\rm \log g$}
\def\al{{ et al~}}
\def\ew {equivalent width}
\def\ews {equivalent widths}
\def\feh{[Fe/H]~}

\def\cogs{curves of growth}
%
% The actual paper
%
\cleartabnumber
%
%\refereelayout
\MAINTITLE{New beryllium observations  in low--metallicity stars
\FOOTNOTE{Based on observations collected at the European
Southern Observatory, La Silla, Chile}},
\AUTHOR={P. Molaro@1, P. Bonifacio@1, F.  Castelli@2@,@1, L. Pasquini@3}
\INSTITUTE={
@1Osservatorio Astronomico di Trieste
Via G.B. Tiepolo 11,  34131 Trieste, Italy
@2 Consiglio Nazionale delle Ricerche, -- 
G.N.A. Via G.B. Tiepolo 11,  34131 Trieste, Italy
@3 European Southern Observatory, Casilla 19001, Santiago 19, Chile
}
\ABSTRACT={
We present observations of the Be II 313.0 nm resonance doublet in 14 halo
and old disk stars  with 
metallicities ranging from [Fe/H]=-0.4 to $\approx$-3.0
obtained with the CASPEC spectrograph of the ESO 3.6m telescope
at   a  FWHM$\approx$ 8.6 \kms ~   resolution.
Abundances are derived by means of the synthetic spectra technique employing 
Kurucz (1993) atmospheric models, with  enhanced $\alpha$-elements and 
no overshooting. 
The derived abundances together with those available in  literature
show that  for -2.7 $<$ [Fe/H] $<$ -0.8  Be correlates linearly with 
iron [Be]$\propto$ 1.07($\pm 0.08$)[Fe/H], giving  strength to previous results. However, a steeper correlation 
is still possible at  metallicities lower than [Fe/H]$<$-1.4 with
[Be]$\propto$ 1.6($\pm 0.44$)[Fe/H]. 
When iron is replaced with oxygen, Be is found tracking closely
oxygen  
up to solar values, without signs of breaking in 
correspondence of the onset of the  Galactic disk.

No evidence of intrinsic dispersion   is found, 
ought to the large errors involved in the Be abundance determinations, but
for three stars (HD 106516, HD 3795, HD 211998) 
a significant upper limit in the Be abundance
 can be placed at  $\approx$ 1 dex  below the mean trend of the Be-Fe relation. For such stars
non conventional mixing is required to explain Be depletion.

 Be observations can be used  to discriminate 
strongly Li-depleted stars. These are the stars which show less Li than
that expected by high energy cosmic 
rays production as deduced from Be observations. The available Be observations 
imply that   some of the stars which contribute to the scatter 
in the Li-Fe  diagramme are  Li-depleted stars. This result 
strongly supports the use of the upper envelope of the Li-Fe diagramme  
to  trace the  Li galactic evolution,  and argues for   a low value 
for  
the  primordial  Li
against  models which predict 
substantial Li depletion in halo and old disk stars.
}
  \KEYWORDS={ Stars: abundances, Stars: Population II, Galaxy: evolution
Cosmology: observations, ISM: cosmic rays}
  \THESAURUS={08.01.1, 08.16.3, 10.05.1, 12.03.3, 09.03.2}
\OFFPRINTS={ P. Molaro, Osservatorio Astronomico di Trieste,
Via G.B. Tiepolo 11,  34131 Trieste, Italy}
\DATE={ ????? }           % must be given at the last moment
\maketitle
\titlea{Introduction}
\par
The importance of  beryllium 
stems largely from its unique origin. Be 
is probably 
a pure product of the galactic cosmic ray (GCR) nucleosynthesis, generated
as the other light elements $^{6,7}$Li and $^{10,11}$B 
through 
the bombardment of $^{12}$C and $^{16}$O in the interstellar medium 
by energetic cosmic rays (Reeves, Fowler and Hoyle 1970;
Meneguzzi et al 1971, Walker et al 1985).
 The process of $\nu$-induced nucleosynthesis
may produce significant  amounts for  
$^7$Li, $^{10,11}$B but only small amounts of $^9$Be 
(Woosley and Weaver 1995).
Standard Big Bang nucleosynthesis predicts negligible quantities of Be out
of the primordial nucleosynthesis, namely  Be/H$\approx$10$^{-17}$ 
(Thomas et al 1994), but
it  has been  suggested that baryon inhomogeneous Big Bang
 nucleosynthesis might produce
enhanced quantities of Be up to four orders of magnitude  
relatively to the homogeneous case
(Boyd and Kajino 1989; Malaney
and Fowler 1989; Kajino and Boyd 1990). However, such  inhomogeneities from
the  quark-hadron phase transition do not appear   very probable  
(Kurki-Suonio et al 1990; Terasawa  and Sato 1990;
Mathews et al 1990; Jedamzik, Fuller and Mathews 1994; Thomas et al 1994).
A primordial production of Be would appear  as
 a {\it plateau} of the Be abundance 
in stars of  low 
metallicities with  analogy to  what is observed for Li, and in a  Be/B ratio
significantly higher than that predicted by the spallation processes.
However, the presence of such a  {\it plateau} 
at low metallicities  is not  an unambiguous primordial signature, 
since it can be also  produced by accretion onto the stellar surface while
the star is
passing  through thick interstellar clouds, as shown by Yoshii, Matwhews 
and Kajino (1995).

Beryllium was observed in the sun by Chmielewski et al.
(1975) at a level of [Be]=1.15 (i.e
Be/H=1.4$\cdot$ 10$^{-11}$); here and after we will follow the notation 
[Be]=$\log$ (Be/H) + 12. In a sample of young hot stars
 Boesgaard (1976) measured almost the same abundance 
of Be, namely $<$ [Be] $>$ = 1.1$\pm$ 0.1.  
The meteoritic value from 
carbonaceous chondrites is presumably the initial solar value and
is [Be]=1.42$\pm 0.04$, which is   about a factor of two
higher than the presently observed
solar value and also a factor of two higher 
than the mean of the stellar values (Grevesse
and Noels 1993).
Observations of Be in halo dwarfs began in the eighties with 
 Molaro and Beckman (1984), Molaro, Beckman and Castelli (1984), 
Molaro (1987), 
showing that Be abundance in the past was lower than the solar one,
which was not obvious at that time.
 Only recently has the technology 
become adequate 
for providing the required sensitivity and resolution at 313.0 nm
and therefore 
accurate observations of the BeII lines became possible. 
The Be abundances in halo stars 
by Rebolo et al (1988), Ryan et al (1991), 
Ryan et al (1992), Gilmore et al (1991), Gilmore et al 
(1992),  Boesgaard and King 
(1993) revealed a clear    linear
correlation with iron, 
while a steeper relation was expected. To account for this  behaviour  
 it has been suggested
that  spallation processes occur 
preferentially close to supernovae (Feltzing and
Gustafsson 1994; Olive et al 1994; Tayler 1995).
A preliminary analysis of the present data was already given in 
Molaro et al (1995a) and Primas (1995).

In this paper we present new observations and Be abundances for 14 halo
and old disk stars. These new data, together with the data available from the
literature, are used to investigate the evolution of Be.

\titlea{Observations and data reduction}

We used the CASPEC spectrograph at the ESO 3.6 m
telescope to obtain ultraviolet spectra of our program stars.
A  log of observations and the available photometric data 
for the stars are given in Table 1.
CASPEC is not normally used at these short wavelengths and
a special instrument configuration had to be devised. 
To enhance resolution we used
the long  camera coupled with  the blue cross-disperser.
During observations 
alignment of the slit with the parallactic angle was 
maintained by rotating the adaptor to avoid  slit losses
due to the atmospheric dispersion.
In order to minimize the red leak a UV filter was used in the
optical path. A serious problem for this configuration
is the acquisition of suitable calibration frames;
in fact the light from the calibration lamps is fed into the
spectrograph through a prism which is opaque in the UV, and 
hence no flat--field or Th--Ar can be obtained in the conventional
way. The Th-Ar spectrum taken in this way is usable only longward of
about 330.0 nm, and therefore  we used an external Th-Ar
lamp. This is mounted in the focal plane of the telescope
and illuminates  the spectrograph slit directly. Spectra of this
``external'' lamp had to be taken at the beginning and at end
of each night, since the operation of mounting and dismounting the
lamp, though simple, is time-consuming. 
The resolving power as measured from the
emission lines of the 
Th-Ar lamp around the Be line region is of $\approx$ 35000.
\par
A similar approach could be used for flat fields, but,
at present,  a suitable UV flood-lamp is not available at ESO.
The analysis of the first images acquired for this program convinced us that 
we could  perform a meaningful reduction without flat--fielding. There is no
fringing and the abundance analysis is done on normalized
spectra, thus making  the
removal of the blaze effect not strictly necessary.
\par
The spectra were reduced using the Echelle package in MIDAS. The major
source of uncertainty in the calibration procedure is the background
subtraction; in fact in the spectral region under study
the orders are closely spaced and the inter--order region 
 contains 
light coming from the tails of the adjacent orders. 
The background
was  taken to be the lower  envelope of the signal in the
inter--order region.  We estimate the uncertainty of the background 
evaluation to be of the order of 10--15\% of the background, 
from the RMS of 
in the inter--order, for most of our spectra, it can
be higher for the low S/N spectra (e.g. HD 128279).
This uncertainty introduces an error of the order of 10\%
in the level of the continuum. For instance in HD 3795
the average background per pixel 
is of 33 ADU, while the continuum is at  61 ADU,
the RMS of the background is of 2.4 ADU, which is 8.6\%
of the continuum level.
The effect of this error on the \ew of Be II lines is discussed in Sect. 5.1.

%
% anchor for the table number
%
\begtabfullwid
\tbl{Log of Observations and photometric data. 
Johnson photometry has been taken from SIMBAD, Str\"omgren photometry
from Hauck \& Mermilliod (1990) and from Schuster \& Nissen (1988)
(in the second row, when available)}

\halign to \hsize{ 
\tabskip=20ptplus60ptminus10pt
\hfill #&
\hfill #&
\hfill #&
\hfill #&
\hfill #&
\hfill #&
\hfill #&
\hfill #&
\hfill #&
\hfill #&
\hfill#\cr
HD\hfill & V &$\rm (B-V)$ & $\rm (b-y)$ & m1 & c1  &$\beta$&
$\rm E(b-y)$
& date & exp (sec) \cr
\cr
\multispan{10}{\hrulefill}\cr
\cr
3795   & 6.14 & 0.70 & 0.450 & 0.213 & 0.304 & 2.565 & 0.017 
& 27-9-93 & 3000 &\cr
6434   & 7.72 & 0.61 & 0.385 & 0.160 & 0.273 & 2.581 & 0.000 
& 31-7-93 & 6300 &\cr
&&& 0.386 & 0.160 & 0.272 & 2.573 & 0.000 \cr
16784  & 8.02 & 0.56 & 0.378 & 0.142 & 0.293 & 2.569 & 0.000
& 28-8-93 & 3600 &\cr
&&& 0.377 & 0.141 & 0.291 & 2.581 & 0.000 \cr
25704  & 8.10 & 0.55 & 0.369 & 0.120 & 0.272 & 2.584 & 0.000
& 28-8-93 & 6600 &\cr
&&& 0.371 & 0.118 & 0.275 & 2.570 & 0.000 \cr
76932  & 5.86 & 0.52 & 0.359 & 0.120 & 0.293 & 2.578 & 0.000
& 13-3-92 & 7200 &\cr
&&& 0.354 & 0.117 & 0.297 & 2.574 & 0.000 \cr
106516 & 6.11 & 0.45 & 0.318 & 0.115 & 0.332 & 2.613 & 0.000
& 14-3-92 & 7200 &\cr
&&& 0.318 & 0.110 & 0.335 & 2.606 & 0.000 \cr
128279 & 7.97 & 0.64 & 0.470 & 0.058 & 0.264 & 2.545 & 0.000
& 30-7-93 & 7200 &\cr
140283 & 7.24 & 0.49 & 0.378 & 0.040 & 0.302 & 2.584 & 0.043
& 14-3-92 & 14400 &\cr
&&& 0.380 & 0.033 & 0.284 & 2.564 & 0.020 \cr
160617 & 8.73 & 0.46 & 0.344 & 0.053 & 0.353 & 2.607 & 0.041 
& 15-3-92 & 7200 &\cr
&&& 0.347 & 0.051 & 0.331 & 2.584 & 0.015 \cr
166913 & 8.23 & 0.45 & 0.327 & 0.078 & 0.314 & 2.610 & 0.010 
& 29-7-93 & 7200 &\cr
200654 & 9.11 & 0.64 & 0.460 & 0.029 & 0.274 & 2.534 & 0.000
& 28-8-93 & 6600 &\cr
&&& 0.460 & 0.027 & 0.271 & 2.534 & 0.000 \cr
211998 & 5.29 & 0.68 & 0.447 & 0.116 & 0.240 & 2.543 & 0.000
& 28-8-93 & 3600 &\cr
&&& 0.450 & 0.109 & 0.249 & 2.546 & 0.000 \cr
218502 & 8.50 &&&&&&& 29-7-93 & 7200 &\cr
219617 & 8.16 & 0.47 & 0.343 & 0.078 & 0.231 & 2.586 & 0.000
& 29-7-93 & 7200 &\cr
&&& 0.349 & 0.072 & 0.243 & 2.584 & 0.000 \cr
\multispan{10}{\hrulefill}\cr
}
\endtab
%
% anchor for the table number
%
\newcount\obs
\obs=\tabnumber

\titlea{Model atmospheres}

For the present  investigation we used version 9 of the ATLAS code to compute
the atmospheric  models appropriate for metal--poor stars.
This version of  the ATLAS code (Kurucz 1993)
differs from previous ATLAS versions mostly for the opacity and for the
way in which the mixing-length convection is handled. In the
ATLAS9 models the opacity distribution functions (ODFs), which account for
the line opacity, were computed with
a much larger number of atomic lines than in the previous versions
and, for cool stars, molecular lines were also
added. Continous opacities were implemented by taking into account
also the contribution of the
OH and CH molecules. Convection is still based on the mixing-length approach,
but two modifications have been made in ATLAS9; the first one allows for a
horizontally averaged opacity and the second one allows for an approximate
overshooting (Castelli 1996).

As far as  convection is concerned, Castelli, Gratton and
Kurucz (1996) have shown that
the first modification
of the mixing-length has negligible effects on the
results, while the second one alters the whole structure of the models,
mostly when  \t~  is  between 5500 K and 8000 K.
Figure 1 compares the T--log~$\tau_{Ross}$ relations for \t~ = 5750 K,
log g=3.5, and [M/H]=-1.0, with and without overshooting. 
The zone of formation of the Be II lines is around
log($\tau_{Ross}=-0.12$ where the
temperature is higher for models with overshooting. 
While
the different convection affects the whole structure,
 the different chemical composition due to
the enhancement of the $\alpha$-elements, 
mostly affects  the structure of the deepest layers, as can be seen from the
figure.
The discrepancy between models with and without overshooting
 changes with log g,  \t and metallicity. 
The Kurucz solar model computed with overshooting
fits  the observations very well, but for other stars
no overshooting models generally yield more consistent \t~  values
when different methods are used to derive them, such as methods based on colors,
Balmer profiles, and the infrared flux.  A test on Procyon
has also shown that, for this star, no overshooting models give parameters
which are more consistent
with the fundamental values, derived from model independent
methods.
\par
On the basis of these results and also because the {\it ad hoc} inclusion of
the overshoot in the ATLAS9 code is not really the physical overshoot
(Freytag 1996), we decided to use for this investigation
ATLAS9 models with  the overshooting option switched off. We kept the
same value l/H$_{p}$=1.25 adopted by Kurucz for the mixing-length to pressure
height scale ratio, because we found that it well reproduces the solar
irradiance also when overshooting is dropped.
 The choice of the mixing length in
the range from 0.5 to 2.0 for the pressure  scale height
makes negligible differences (less than 0.01 dex) over the whole grid, and
the adopted microturbulent velocity does not
affect the derived Be abundance  for the range of equivalent width 
under consideration. 
\par
 We used the $\alpha$--enhanced opacity distribution
functions (ODFs)
provided by Kurucz (1993a), which are obtained  by assuming that
all  $\alpha$~ elements are enhanced by 0.4 dex over iron.
 This is certainly
a more realistic 
chemical composition for 
Pop II stars than the usual approach which uses   the ODFs with solar-scaled
abundances.  
Furthermore, for these ODFs the solar iron abundances is
log (N$_{Fe}$/N$_{tot}$)=-4.53 (Hannaford et al 1992), instead of
log (N$_{Fe}$/N$_{tot}$)=-4.37 (Anders and Grevesse 1989), which was used
for all the ODF's for solar and solar-scaled abundances.
\par
We selected the ODFs 
computed with a microturbulent velocity
of 1 \kms  to allow for very low microturbulent
velocities, as can be  found in some Pop II stars, instead of the 2 \kms
of the  Kurucz (1993a) grid.
The grid covers the range
from 4750 K to 6250 K in \t~ at steps of 250 K, from 2.50 to 4.75 in \g~
at steps of 0.25 and from -0.5  to -3.00 in \feh at steps of 0.5.
For all the models of this grid we
also computed the emergent flux and the Johnson UBV and
the Str\"omgren {\it uvby} colour indices. Colour indices were then used to
derive  effective temperature and surface gravity
for all the stars of our sample, except HD 218502 for which no observed
photometric Johnson or Str\"omgren indices
have been found in the literature.
Model parameters for the stars of our sample will be discussed
in the next section.
\begfig 6.5 cm
\figure{1}{
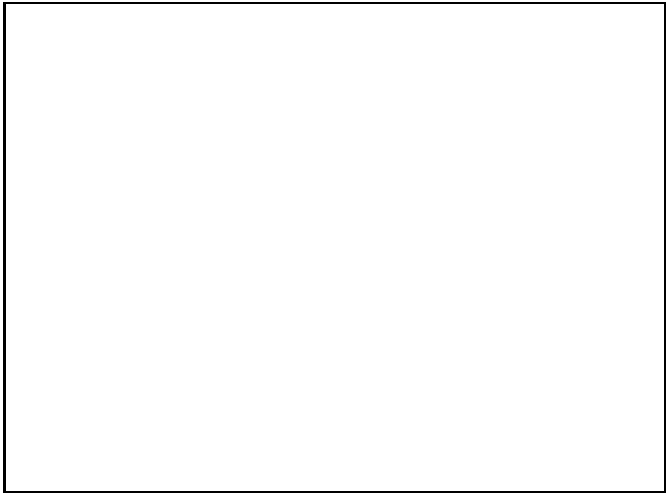
Temperature structure for models with $T_{\rm eff}$ =5750,
$\log g$ = 3.50 and metallicity [Fe/H]=-1.0: $\alpha$ 
enhanced with (x) and
without (squares) overshooting;
no $\alpha$ enhancement with(+) and without (triangles) overshooting
}
\endfig

\par
In order to appreciate the
differences with the Kurucz (1993a) grid
we computed models for \t = 5250, 5750, 6250 K,
\g = 3.0, 3.5, 4.0 and metallicity [Fe/H]= -1.0, -1.5, -2.5
with the $\alpha$ enhanced ODFs and  
 the overshooting option, and other models  with the
same parameters, but using the ODFs without $\alpha$
enhancement and   overshooting.
\par
\begfig 12 cm
\figure{2}{
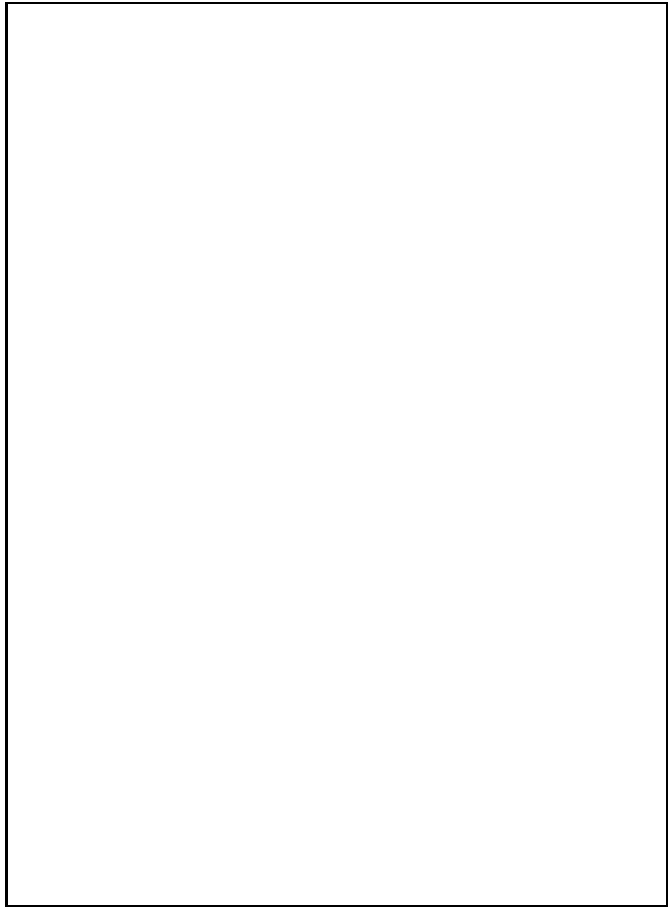
Curves of growth for the BeII 313.1065 nm 
line for models with $T_{\rm eff}$ =5750,
log g = 3.50 and metallicity [Fe/H]=-1.0: $\alpha$ enhanced with (x) and
without (squares) overshooting; 
no $\alpha$ enhancement with(+) and without (triangles) overshooting
}
\endfig
For all these models, as well as for those from our grid and
from the Kurucz(1993a) grid, we computed  the \cogs~ 
for the Be II 313.1065 nm
  line. 
Figure 2 shows the different curves of growth for  Be II 313.1065 nm
corresponding to  the same model parameters \t~ = 5750 K, log g=3.5,
and [M/H]=-1.0 and to the different T-log~$\tau_{Ross}$
relations displayed in figure 1. 
 \par
 In the very low metallicity domain, i.e.
\feh $<$ -2.0,  the effect of $\alpha$~ element enhancement
is very small, of the order of 0.01 dex at most, with
the $\alpha$~ enhanced models yielding  higher abundances.
 On the other hand,
the effect of overshooting is not negligible and   the
models with overshooting yield abundances which are higher by as
much as 0.08 dex. This difference   is almost constant at all temperatures.

 In the low metallicity domain
(i.e. \feh between
-2.0 and -1.0 )  the effect of overshooting is
of the same order of magnitude as in the very low metallicity regime,
but the effect of $\alpha$~ enhancement begins to be  notable in particular
for the cool models. Models with enhanced $\alpha$-elements  yield
 higher abundances of the order of 0.08 dex at \t = 5250 K , but of only 0.03
dex at \t = 6000 K and less than 0.01 dex at \t =6250 K.
No overshooting in the models implies lower abundances, while
$\alpha$~ enhancement implies higher abundances so that 
there is some compensation between the two effects. 
The  result of these opposite behaviours is that
   while  our models yield abundances  larger
than those of the Kurucz 1993a grid at 
low-temperatures, the reverse is true at
the high temperature end.

For solar metallicities there is little or no difference in the Be 
abundances with respect to the implementation of overshooting in the models.
 This is because metal-poor models have convective zones which
start at much shallower depths (D'Antona and Mazzitelli 1984).

Our models yield abundances which are identical within few hundredths of a dex
 to  those  obtained using Kurucz 1979 models and are therefore
directly comparable with  the Be  literature values which are obtained
mostly using those models. This  also implies  that the much larger
number of lines included in the computations of the ODFs implemented in
version 9 of the ATLAS code has relatively small effects on the
temperature structure of the models.
We stress 
that     
our choice of switching off overshooting makes the present analysis
consistent with the older grids of models which make
use of either Gustafsson-Bell or old (i.e. computed with version 8
or earlier ones of the ATLAS code)
Kurucz models.

 In conclusion,  the 
differences in Be abundances found at low metallicities when using the 
Kurucz 1993a models,
 with respect to the Kurucz 1979 ones, are due
to the presence of overshooting.
The low effect  on the metal-poor stars of the increased blanketing
in the new models is explained by the small importance of the new lines in the
metal poor stars. In fact, the large number of atomic lines  added
for computing the new ODFs  arise from high-excited states and
are therefore weak lines in solar metallicity stars. The effect of the
molecular lines has still to be investigated.
The abundances derived from lines whose depth of formation is
close to the top of the convection zone, as  for Be II and also for LiI,
are  quite sensitive to the assumptions made on overshooting
(Molaro et al 1995b,c).
\par
This extensive theoretical grid  allowed us to perform
an efficient  comparison between observed and computed quantities
(spectra and colours) and also to experiment the effects of
small changes in the atmospheric parameters on the derived
abundances.
In addition to the models of the grid, more models
were computed with atmospheric parameters appropriate
for our program stars (see next section), including
a small number of models with \feh = -0.5
for which we used ODFs with solar--scaled abundances rather
than with $\alpha$~ enhanced, the other assumptions being the
same as  the more metal-weak models.

\par
 Non-LTE effects have been shown to be negligible in the sun by
Chmielewski et al (1975). For metal-poor stars absolute NLTE
corrections to the Be abundances have been shown to be lower than
0.1 dex by Garcia Lopez et al (1995) and Kiselman \& Carlsson (1995).

\titlea{Atmospheric parameters}

%
% anchor for the table number
%
\newcount\TB
\begtabfullwid
\tbl{Atmospheric parameters  for the program stars}
\TB=\tabnumber
\def\s{\phantom{$^c$}}

\halign to\hsize{
\tabskip=20 pt plus 10pt minus 40pt
\hfill #&
#\hfill &
#\hfill &
\hfill #&
\hfill #&
\hfill #&
\hfill #&
\hfill #&
\hfill #&
\hfill #\phantom{1}\cr
\multispan{10}{\hrulefill}
\cr\cr
HD &  $T_{\rm eff}$ $^1$& $T_{\rm eff}$  $^2$ 
& log g $^3$ & log g$^4$ & log g$^5$ & log g$^6$   
& [Fe/H]$^7$\hfill & [Fe/H]$^8$\hfill & [Fe/H]$^9$\hfill\cr
\multispan{10}{\hrulefill}
\cr\cr
3795   & 5464 & 5420$^a$ \s       & 4.05 & 4.00 & 3.63$\pm 0.24$ 
       &  & -0.25 &   -0.73$^a$ &  \cr
6434   & 5623 & 5671$\pm 79$ \s  & 4.21 & 4.25 & 4.25$\pm 0.22$
& 4.08$\pm 0.14$  & -0.54 & -0.54$^b$  &-0.68\cr
16784  & 5747 & 5564$\pm 71$ \s   & 4.00 & 3.88 & 3.80$\pm 0.22$
& 3.68$\pm 0.13$ & -0.67 & -0.75$^c$  &-0.63\cr
25704  & 5829 & 5884$^b$  & 4.19 & 4.00 & 4.23$\pm 0.20$
&  & -0.93 & -0.85$^b$ &\cr
76932  & 5932 & 5965$^b$ & 4.15 & 4.00 & 4.18$\pm 0.17$
&  & -0.78 & -0.82$^b$ &\cr
106516 & 6211 & 5995$\pm 68$ \s   & 4.23 & 3.98 &  4.00$\pm 0.14$
& 3.97 $\pm 0.13$ &  -0.75 & -0.70$^d$  &-0.86\cr
128279 & 5200 & $5165\pm 77$ \s   & 2.95 & 2.95 & 2.88$\pm 0.14$
& 2.98 
$\pm 0.13$ & -2.22 &  -2.50$^e$ &-1.97\cr
140283 & 5864 & 5814$\pm 44$ \s   & 3.66 & 3.58 & 3.48$\pm 0.10$
& 3.27 $\pm 0.09$ & -2.00 & -2.75$^f$      &-2.36\cr
160617 & 6144 & 5664$\pm 84$ \s   & 3.82 & 3.13 & 3.08$\pm 0.16$ 
&3.29 $\pm 0.23$
& -1.80 &  -2.05$^f$      &-1.76\cr
166913 & 6213 & 5955$\pm 109$ \s  & 4.16 & 3.85 & 3.73$\pm 0.18$ 
& 4.45 $\pm 0.19$ & -1.32 & -1.80$^g$ &-1.31\cr
200654 & 5208 & 5522$\pm 119$ \s  & 2.78 & 3.15 & 3.13$\pm 0.17 $
& 3.56 $\pm 0.21$ & -2.65 
& -3.13$^f$ &     -2.38\cr
211998 & 5243 & 5338$\pm 65$ \s  & 3.57 & 3.65 & 3.55$\pm 0.26$
& 3.26 $\pm 0.12$ & -1.43
& -1.54$^h$     &-1.40\cr
218502 & & 6000$^i$ &&&& 3.80$^i\phantom{\pm 0.10}$ 
&    & -1.96$^l$    &\cr
219617 & 5960 & 5815$\pm 76$ \s      & 4.36 & 4.23 & 4.05
$\pm 0.18$ & 3.44 $\pm 0.14$
& -1.32  & -1.63$^h$      &-1.08\cr
\cr
\multispan{10}{\hrulefill}\cr
}
\vskip 1 cm\noindent
$^1$ from $\rm (B-V)$; $^2$ from Balmer lines, Fuhrmann et al (1993);
$^3$ from c1 photometric temperatures ($T_{\rm eff}$ $^1$) and
photometric metallicities ([Fe/H]$^4$);
$^4$ from c1 spectroscopic temperatures ($T_{\rm eff}$ $^1$)
and photometric metallicities ([Fe/H]$^7$);
$^5$ from c1 spectroscopic temperatures ($T_{\rm eff}$ $^1$)
and spectroscopic metallicities ([Fe/H]$^8$);
$^6$ spectroscopic from Axer et al (1994) with Teff of column 2 and
metallicities of column 8;
$^7$ photometric metallicities;
$^8$ spectroscopic metallicities from literature;
$^9$ spectroscopic metallicities from Axer et al (1994) with the 
Teff of column 3.
$^a$ Pasquini et al (1994),
$^b$ Edvardsson et al (1993),
$^c$ Ryan \& Norris (1991),
$^d$ Spite et al (1994),
$^e$ Peterson et al (1990),
$^f$ Nissen et al (1994),
$^g$ Magain (1989),
$^h$ Gratton (1989),
$^i$ Luck \& Bond (1985)
$^l$ Pilachowski et al (1993)
\endtab

The  appropriate
model--atmosphere for  the sample stars  
 is  specified by \t, \g~ and metallicity.

Ten stars out of our  14 star sample  have 
been studied by Fuhrmann et al (1994)
who   determined \t~ from the Balmer profiles
of a large sample of G--dwarfs. 
Surface gravities and spectroscopic 
metallicities from Axer et al (1994) are given in column 7 and 10 of
Table 2. 
The Axer et al  
gravities and metallicities are spectroscopically derived  assuming the
Fuhrmann et al (1994) temperature, thus forming a set of homogeneously
derived stellar parameters.
\par
A photometric estimate of the effective temperature may be derived
from the
$\rm (B-V)$
 index, which is almost independent of gravity between 5750 K and 6250 K.
For each star we interpolated our UBV grids for 
an assumed metallicity and gravity
to obtain the \t corresponding to the observed $\rm (B-V)$.
For the reddened stars, we dereddened the observed $\rm (B-V)$ indices by using
the relation $\rm E(B-V)=E(b-y)/0.74$ from Crawford \& Mandwewala (1976),
and the $\rm E(b-y)$ value derived below.
These photometric temperatures may be found in column 2 of Table 2.
\par
For a sample of seven subdwarfs,  Castelli, Gratton and Kurucz (1996)
showed that \t~ derived from H$_{\alpha}$ profiles computed with
models similar to those used in this paper (namely ATLAS9 models with
the overshooting option switched off) may differ from 10 K to 200 K from
those derived by Axer et al. (1994).
The differences between the photometric determinations of \t~  based
on our models and the effective temperatures
adopted for computing the Be abundances are of the same order,
except for  HD~160617, HD~200654, HD~166913, and HD~106516 for which
the differences are 480 K, 314 K, 258 K, and 216 K respectively.
For HD~160617 and HD~166913 Magain (1989) derived \t~ equal to 5910 K and
6030 K respectively. The above differences would reduce to 234 K and 183 K
respectively.

\par
We derived also a photometric
  surface gravity  for
 the stars with  available Str\"omgren photometry. 
The synthetic Str\"omgren indices have been computed from our grid of fluxes
with $\alpha$ enhancement and no overshooting, as described in the previous
section.
The use of Kurucz (1993) synthetic colours  yields  log g
being generally  0.1 to 0.2 dex lower than our estimates.
\par
The observed 
c1 indices  were taken from  the electronic version
of the Hauck \& Mermilliod (1990) catalogue supported at CDS and 
corrected for reddening using the intrinsic colour
calibration of Schuster \& Nissen (1989), which is most appropriate
for late-F and G-type metal-poor stars. For each star we iterated
their equation (1) until the change in $\rm (b-y)_0$ was less than
0.0001 mag.  The derived reddening is always very small, as expected since
our stars are nearby, yet non--zero in a few cases.
For those stars for which 
\eby ~ has small negative values we imposed \eby = 0.000. Also for the
stars with $\beta$$<$ 2.55, which lie outside the range of the calibration,
we assumed zero reddening.
\par
The surface gravities were derived by comparing the theoretical and
observed c0 for an assumed effective temperature and metallicity.
 For the
reddened stars c0 was dereddened by means of the relation
$\rm E(c)=0.2E(b-y)$ (Crawford 1975).

Column 4 of Table 2 gives the surface gravities derived from the
photometric temperatures of column 2 and photometric metallicities
of column 8. In columns 5 and 6 the Fuhrmann et al temperature
is assumed instead, but while in column 5 the photometric
metallicity of column 8 is assumed, in column 6 
the ``best'' literature value of column 9 is taken. To be brief, errors on the
photometrically derived log g are given only in column 6,
the others being comparable.
\par

We derived 
photometric metallicities for the stars with available  Str\"omgren photometry
(i.e. all except HD 218502) derived from
 Schuster \& Nissen's
(1989)
calibration. The values are given in column 8 of Table 2.

The result of a search  for spectroscopic determinations
of the metallicity in literature  is reported   in column 9
of Table 2. 
It is worth noting that when  comparing the 3 different sources for
metallicities only in a few  cases is  the disagreement 
larger than 0.5 dex: namely HD 128279,
HD 166913, HD 200654 and HD 219617.

A direct check of the gravity is possible for HD 140283
which has a measured parallax from where Nissen \al (1994)
estimate \g = 3.39$\pm 0.15$; this value is consistent
with all  four values reported in Table 2. 
On the other hand, remarkable
disagreements up to about 0.7 
dex are found between the spectroscopic and photometric
gravities for  HD 166913 and  HD 219617.
However, this is not so uncommon, for instance,
for HD 76932 the gravities in the literature
show a very large spread ranging from the  3.5 of Bessel \al
(1991) to 4.37 of Edvardsson \al (1993).
For HD 166913 
there is  a considerable uncertainty in the atmospheric
parameters. Laird (1985) gives \t = 6120 and log g = 4.43.
Fuhrmann et al and Axer et al give \t = 5955 and  
\g = 4.45, whereas from the photometry we deduce a log g=3.8.
This uncertainty in 
the atmospheric parameters  represents  a major source of uncertainty in the Be abundances. 
\par

For our analysis we  adopted the
effective temperature provided by
Fuhrmann \al (1994) and the surface gravity
of Axer et al (1994) whenever  available. For these stars
the metallicity of the
adopted model was that of Axer et al (1994), rounded to the nearest 0.5 dex. 
 For HD 3795, HD 25704 and
HD 76932, which
are not included in the paper of Fhurmann et al (1993), we adopted
the photometric gravity derived by using the literature metallicity reported
in  column 6 in Table 2.
For these stars 
the effective temperature 
has been taken on the basis of a critical analysis of
the literature. These temperatures are close to our photometric estimates. 
 For HD 218502, which lacks the necessary photometry,
we took gravity and \t~ from Luck \& Bond (1985).

\begtabfull
\tbl{Berylium abundances}
\halign to \hsize{ 
\tabskip=20ptplus60ptminus10pt
\hfill #&
\hfill #&
\hfill #&
\hfill #&
\hfill#\cr
HD\hfill & Teff& Log g & [Fe/H] & [Be]\hfill \cr
\cr
\multispan5{\hrulefill}\cr
\cr
3795   & 5420 & 3.60 & -0.73  &$< -0.48   $\cr
6434   & 5671 & 4.08 & -0.68  &$ 0.81 \pm 0.22  $\cr
16784  & 5564 & 3.68 &  -0.63 &$ 0.19  \pm 0.21 $\cr
25704  & 5884 & 4.20 &  -0.85 &$ 0.39 \pm 0.24 $\cr
76932  & 5965 & 4.15 &  -0.82 &$ 0.79 \pm 0.21 $\cr
106516 & 5995 & 3.97 &  -0.86 &$< -0.76   $\cr
128279 & 5165 & 2.98 &  -1.97 &$ -0.75 \pm 0.30 $\cr
140283 & 5814 & 3.27 &   -2.36&$ -0.91 \pm 0.17 $\cr
160617 & 5664 & 3.29 &  -1.76 &$ -0.90 \pm 0.27 $\cr
166913 & 5955 & 4.45 &  -1.31 &$ 0.54  \pm 0.15$\cr
       &      & 3.73  &        & $  0.23$$^a$$\pm 0.15$\cr
200654 & 5522 & 3.56 & -2.38  &$ -0.56 \pm 0.22$\cr
211998 & 5338 & 3.26 & -1.40  &$<-1.05     $\cr
218502 &      & 3.80 & -1.96  &$ -0.56  \pm 0.22$\cr
219617 & 5815 & 3.44 & -1.08  &$ -0.56  \pm 0.20 $\cr
       &      &  4.05 &        &$ -0.16$$^b\pm 0.20$\cr
\multispan5{\hrulefill}\cr
}
\vskip 1cm\noindent
$^a$ if log g = 3.73 ; $^b$ if log g = 4.05
\endtab
%
% anchor for the table number
%
\newcount\Be
\Be=\tabnumber

\titlea{Be abundances}
The  beryllium abundances  are
 derived from the Be II $^2S-^2P_0$ resonance lines.
While the Be II 313.1065 nm line is
unblended, the Be II 313.0420 nm line is blended with OH
313.0281 nm,  V II 313.0257 nm and CH 313.0370 nm,
and hence the abundance of these elements affects the appearance of
the whole feature.
We  therefore determined
the Be abundance from the spectral synthesis
of the 312.6 -- 313.4 nm region.
A work in progress is analysing the whole  
310--350 nm range available  which shall, eventually,
lead to the  determination of the abundances of
C, N, O, Ca,
Ti, V, Cr, Zn.

Model atmospheres
 with the  appropriate atmospheric parameters 
were computed for each star as
 described in Sect. 4.
We used the \ew ~ and the model--atmosphere as input
to version 9 of the WIDTH code (Kurucz 1993a) to derive a first
estimate of the Be abundance. This abundance and the model--atmosphere
were then used as input to the SYNTHE code (Kurucz 1993b) to
synthesize the whole region from 312.6 nm to 313.4 nm.
This synthetic spectrum and the observed spectrum, which was
 shifted in wavelength so that the lines appeared at their
laboratory position,
were superimposed on the same plot. These plots were
visually inspected and then  
the abundances of several elements, including Be, were changed in order to
obtain the agreement between observations and computations.
These abundances were used to compute a new synthetic spectrum
and the whole procedure was iterated until the agreement
was judged satisfactory.
We imposed also 
the requirement that  the blended Be II
313.0420 nm line  should be  reproduced with the same abundance of
the Be II 313.1065 line. 
\par
 The results of this analysis are given in Table 3.
 For the stars HD 166913 and HD 219617 for which the difference
 between the photometric and spectroscopic gravities is particularly
 large we derived the abundance for both  gravity values.

\titleb{Errors}

 The process of abundance determination is complex and the
 estimate of the error  is by no
 means straightforward.
Errors in \g , \t and metallicity will affect the derived
abundances. Microturbulence is unimportant since
for our stars the Be II lines are essentially weak lines.
Given the measured \ew ~of the Be II 313.1065 line,
we interpolated  our grid  of \cogs ~ to estimate
the Be abundance for our adopted parameters and
for \g = \g  $\pm\sigma(\rm log~g)$.

The errors in 
photometrically derived surface gravities are reported in Table 4.
To compute the total $\sigma(\rm log g)$
 ~we have assumed an error of 0.008 mag
on c1, the error in \t ~ provided by 
Fuhrmann et al (1994) or $\pm$ 100 K for the stars not included in
that paper, an error of $\pm$ 0.3 dex in metallicity
and an error in the reddening equal to the reddening value,
which is appropriate for slightly reddened stars.
 For all  parameters, in case of
asymmetric errors we assumed 
the largest one in absolute value. 
Column 5 of Table 4 gives the uncertainty in \g ~due to 
random errors in the photometry, column 6 that due to the
uncertainty in reddening, column 7 that due
to the uncertainty in \t and column 8 that due
to the uncertainty in metallicity.
Finally the square root of the sum of the squares of
these four uncertainties was taken as the total
uncertainty on \g ~and given in column 9.
\par
%
% anchor for the table number
% inserisci tabella  4 e 5 qui
\begtabfull
\tbl{Errors on \g}
\halign to \hsize
{        \tabskip=20ptplus30ptminus20pt
\hfill#~~&
\hfill#&
\hfill#&
\hfill#&
\hfill#&
\hfill#&
\hfill#&
\hfill#&
\hfill#\cr
\multispan9{\hrulefill }\cr
\cr
HD\hfill & $\rm T_{eff}$ & [Fe/H]& log g &$\Delta_{phot}$ 
& $\Delta_r$ & $\Delta_T$ &
$\Delta_m$ &$\rm  \Delta log g$\cr
(1)\hfill&(2)\hfill&(3)\hfill&(4)\hfill&(5)\hfill&(6)
\hfill&(7)\hfill&(8)\hfill&\hfill(9)\cr
\multispan9{\hrulefill }\cr
\cr
\cr
 3795&   5420.& -0.73 &  3.63&  .08&   .02&  .10&  .20&  .24\cr
 6434&  5671.&  -0.54 & 4.25&  .07&   .00&  .05&  .20&  .22\cr
16784&   5564.&  -0.75 &  3.80&   .07&    .00&   .07&   .20&   .22\cr
25704&   5884.&   -0.85 & 4.23&   .08&    .00&   .10&   .15&   .20\cr
76932&   5965.&   -0.82 & 4.18&   .05&    .00&   .13&   .13&   .17\cr
106516&   5995.&   -0.72 & 4.00&   .05&    .00&   .10&   .12&   .14\cr
128279&   5165.&  -2.50 & 2.88&   .08&    .07&   .05&   .08&   .14\cr
140283&   5814.&   -2.75 &3.48&   .05&    .02&   .08&   .05&   .10\cr
160617&   5664.&   -2.05 & 3.08&   .05&    .03&   .13&   .07&   .16\cr
166913&   5955.&   -1.80 & 3.73&   .05&    .03&   .15&   .08&   .18\cr
200654&   5522.&   -3.13 & 3.13&   .05&    .03&   .17&   .05&   .17\cr
211998&   5338.&   -1.63 & 3.55&   .10&    .02&   .05&   .23&   .26\cr
219617&   5815.&   -1.70 & 4.05&   .08&    .00&   .10&   .13&   .18\cr
\multispan9{\hrulefill }\cr
 }
 \endtab
\newcount\TG
\TG=\tabnumber

\begtabfull
\tbl{Errors on Be abundances}
\halign to \hsize
{        \tabskip=20ptplus30ptminus10pt
\hfill#~~&
\hfill#&
\hfill#&
\hfill#&
\hfill#&
\hfill#&
\hfill#\cr
\multispan7{\hrulefill }\cr
\cr
HD\hfill &$\delta_g$ &$\delta_T$ &$\delta_m$ &
$\rm\delta [Be]_{sys}$&$\rm\delta [Be]_{stat}$&$\rm\delta [Be]$\cr
(1)\hfill&(2)\hfill&(3)\hfill&(4)
\hfill&(5)\hfill&(6)\hfill&(7)\cr
\cr
\multispan7{\hrulefill }\cr
\cr
3795&       .14&   .03&   .09&   0.17 & 0.15 & 0.32\cr
6434&        .08&   .03&   .08&   0.12 & 0.10 & 0.22\cr
16784&       .07&   .03&   .08&   0.11 & 0.10 & 0.21\cr
25704&       .11&   .04&   .08&   0.14 & 0.10 & 0.24\cr
76932&       .09&   .00&   .07&   0.11 & 0.10 & 0.21\cr
106516&      .06&   .01&   .05&   0.08 & 0.10 & 0.18\cr
128279&      .08&   .02&   .05&   0.10 & 0.20 & 0.30\cr
140283&      .04&   .02&   .02&   0.05 & 0.12 & 0.17\cr
160617&    .10&   .01&   .06&   0.12 & 0.15 & 0.27\cr
166913&      .08&   .01&   .06&   0.10 & 0.05 & 0.15\cr
200654&      .11&   .02&   .04&   0.12 & 0.10 & 0.22\cr
211998&      .06&   .00&   .09&   0.11 & 0.10 & 0.21\cr
218502&      .11&   .03&   .02&   0.12 & 0.10 & 0.22\cr
219617&     .06&   .02&   .05&   0.08 & 0.12 & 0.20\cr
\cr
\multispan7{\hrulefill }\cr
 }
 \endtab
\newcount\TERR
\TERR=\tabnumber

%
% anchor for the table number
%
 
\par
\begfigwid 23 cm
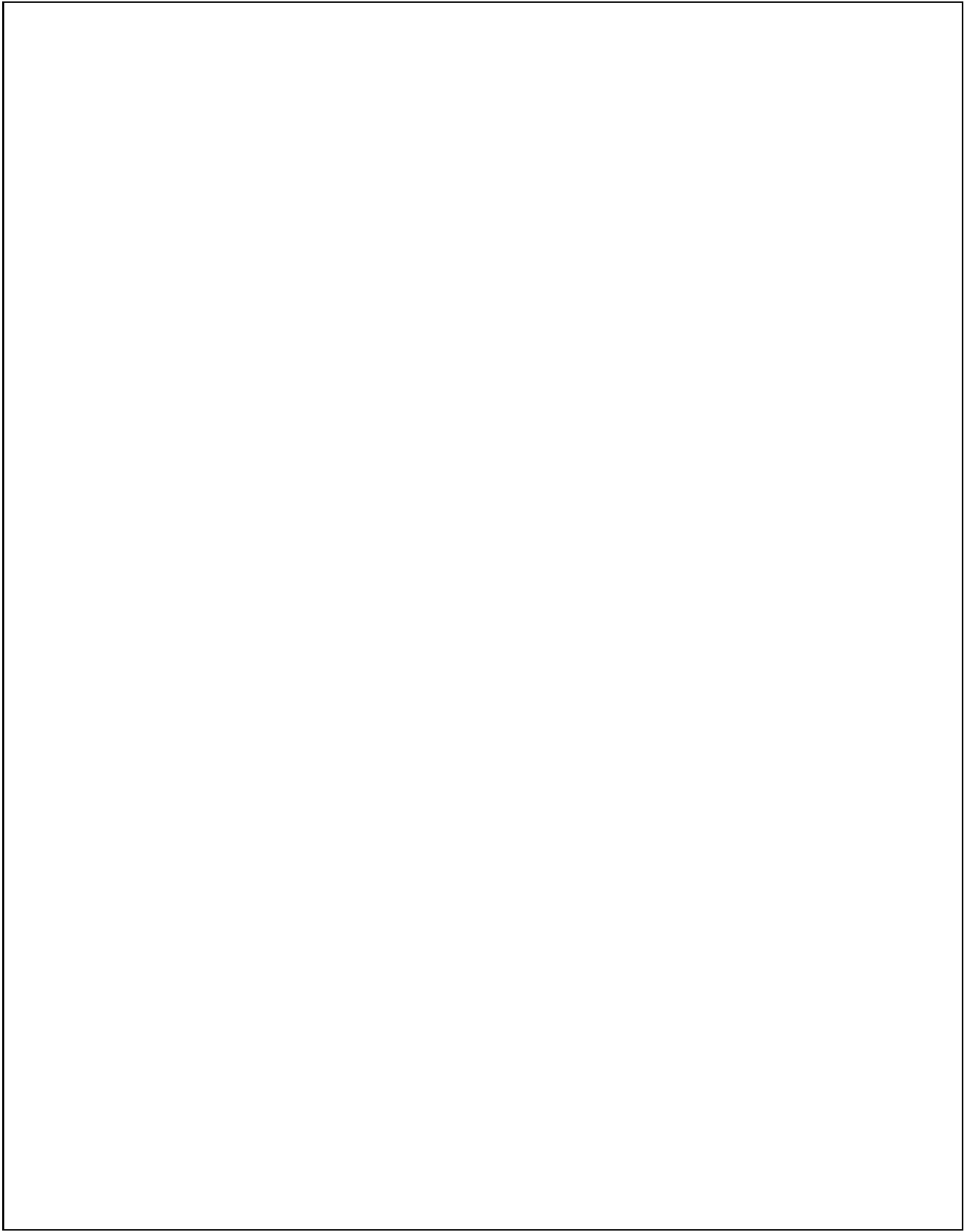
\figure{3}{All the data is displayed, solid lines are 
the observed spectra,
dotted lines are the synthetic spectra, the dashed line above each
spectrum represents the level of the continuum. 
The crosses on the spectra of HD 160617 and HD 211998 indicate cosmic
ray hits which have been removed for display purposes
}
\endfig
\begfigwid 23 cm
\figure{3}{ ~}
\endfig
\par
For the errors in \t and metallicity we proceeded in the
same fashion. 
 For HD 218502, for which \g ~has been taken from the
literature, we arbitrarily assumed an error in \g ~
of $\pm$0.25 dex.
The results and the total error
on [Be] due to errors in the atmospheric parameters
are given in column 5 of Table 5. 
As  expected, the largest error comes
from the uncertainty in \g . These errors ought to
be considered as $1\sigma$ errors. 
\par
 In column 6 of Table 5 
 we give an estimate
 of the error associated with the noise in the data, based
 on several synthetic spectra computed with different
 Be abundances. This uncertainty is summed 
 with the uncertainty associated with the atmospheric
parameters and given in the last column of Table 5.
Both the 
uncertainty in the background subtraction and the
uncertainty in the continuum placement
will affect the equivalent width of the lines.
The effect will be larger, in percentage, for stronger
lines in which wings are important. In our whole  sample
the Be II lines may be considered  weak lines. 
As discussed in Sect. 2, 
uncertainty in the background may lead to an error
 in the continuum level of 10\% for the average quality spectra in our sample.
Trials have shown that this 
will affect the measured \ew ~ by no
more than 0.5 pm . 
On top of this, there may be an error in the placement of the
continuum due to the lack of line--free regions. 
This error
may not be easily estimated for the more metal--rich
stars, while it
can be done for the most metal--poor stars, for which,  with the
aid of synthetic spectra, we may locate regions which are almost
line -- free. In these cases the error on the placement of
the continuum is of the order of 2\%.
Garcia Lopez et al (1995) estimate these errors to be
of the order   of
0.05--0.10 dex in Be abundances for an uncertainty of
2\% in continuum placement. 
Such error  is not included in our estimates.

HD 128279 shows an inconsistency between the two
Be II lines:  the fainter line  is apparently present, while there
is no evidence for the
stronger line. 
Since this is the coolest star of our sample (\t = 5165 K),
and  Garcia Lopez \al (1995) have shown that for stars cooler than
$\approx$ 5200 K the observed feature at the position
of the Be II 313.1 line is  probably contaminated by 
 Mn I $\lambda$ 313.1037 nm, we checked if such a line could be
also responsible for the
feature observed at 313.1 nm in our spectrum of HD 128279. 
In order to explain the solar spectrum Garcia Lopez \al ~ increased
by 1.5 dex the log gf value in Kurucz's line lists.
However, even with this log gf we are not capable
of reproducing the observed feature.

\titlea{The Boesgaard and King sample}

Boesgaard \& King (1993) published a considerably large
data set of Be abundances in halo and old disk stars.
As it   appears from the  discussion given in section 2
our abundances
may not be directly compared to those of Boesgaard \& King (1993),
who employed ATLAS 9 models with overshooting.
There is a further peculiarity in that
their published \cogs ~ yield abundances which are some 0.03 dex
larger than those obtained from the Kurucz 1993a grid.
We suspect that this is due to the fact that BK used a
preliminary version of the 1993 grid, which was slightly
different from the official 1993 release.

Since the models adopted by them are different from
the ones employed in the present study
we have recomputed the Be abundances for all of their halo
stars, where the differences between the models are important,
adopting their original \ews ~ and atmospheric parameters.

The new abundances worked out with models
that account for the $\alpha$-element enhancements and without the
overshooting are given in 
Table 6 and are typically $\approx$ 0.1 
dex lower than originally  estimated by Boesgaard and King (1993).

It is worth noting that in the case of HD 84937, for which $^6$Li  has
been detected by Smith et al (1993), 
the decrease in the Be 
abundance makes a larger $^6$Li/$^9$Be ratio.  The new ratio 
becomes 56, 
which is much larger than the GCR  ratio  $^6$Li/$^9$Be(p+CNO)$\approx$ 5
showing clearly the effect
of $\alpha$-$\alpha$ fusion reactions in the production of $^6$Li.

As discussed in section 3 the models used here are consistent with
both old Kurucz models and Bell-Gustafsson models so that there 
is no need for   revision of the other Be analysis.

\begtabfull
\tbl{Be abundances for BK halo stars}
\halign to\hsize{
\tabskip=40 pt plus 40pt minus 40pt
\hfill #&
\hfill #&
\hfill #&
\hfill #&
\hfill #\cr
HD & $T_{\rm eff}$ & log g\hfill &  \rm [Be] & \rm [Fe/H]\hfill \cr
   &  K\hfill      &[g]=$gcms^{-2}$\hfill & &   \cr
\multispan5{\hrulefill}
\cr\cr
19445  & 5810    & 4.37    & -0.27 & -2.13 \cr
64090  & 5380    & 4.30    &  0.05 & -1.78 \cr
76932  & 5790    & 3.65    &  0.94 & -0.98 \cr
84937  & 6250    & 4.00    & -0.95 & -2.34  \cr
94028  & 5820    & 4.18    &  0.33 & -1.52  \cr
103095 & 5050    & 4.50    &  0.07 & -1.23  \cr
134169 & 5710    & 4.01    &  0.70 & -0.96  \cr
140283 & 5660    & 3.56    & -0.89 & -2.72  \cr
189558 & 5580    & 4.00    & +1.03: & -1.33 \cr
194598 & 5820    & 4.21    &  0.23 & -1.28  \cr
195633 & 5800    & 3.90    &  0.18 & -1.07  \cr
201989 & 5560    & 4.08    &  0.57 & -1.14  \cr
201891 & 5780    & 4.42    &  0.64 & -1.08  \cr
221377 & 6180    & 3.77    & -1.03 & -1.14  \cr
\cr
\multispan5{\hrulefill}\cr
}
\endtab
%
%
% anchor for the table number
%
\newcount\TBK
\TBK=\tabnumber

\titlea{Results}

 Our Be measurements together with
all the Be observations from the literature 
reported in Table 7  are displayed in Fig. 4. 
The data points
are from Rebolo et al (1988), Ryan et al (1991, 1992), 
Gilmore et al (1991, 1992),
Boesgaard and King (1993), Garcia Lopez et al (1995). 
to compute [Be/Fe] we used
the   meteoritic Be value from Grevesse and Noels
(1993). The use of the solar value of   Chmiliewsky et al (1975)
will offset all the data points by +0.3 dex. 
 From Table 7 we may single out the  extreme case  of  HD
189558 for which the difference between the Boesgaard and King (1993) and 
Rebolo et al (1988) determinations   is about 1 dex with almost the same
  stellar
parameters   in both analyses. 
 Also striking is the
case of HD 200654,  since the upper limit of Gilmore et al (1992)
is in sharp  contrast with the detection given here.
The appearance of their spectrum  is different from
the observations shown here (Gilmore private communication), and it may be that
our data were flagged by noise.
Stellar parameters for HD 200654 are also rather uncertain in 
literature. Nissen \al (1994) derive \t = 5090 K, \g =2.7
and [Fe/H] = -3.0, which are  the parameters adopted by
Gilmore \al (1992) to compute the Be abundance. If  so
the star  is a subgiant
with a  \t~ which allows Be dilution.  
 The Fuhrmann \al temperature   is 5522 K,
 significantly hotter, and the gravity from Axer et al (1994) is 3.56,
 so that the dilution is marginal according to 
 standard models.
In Fig. 4 are also  reported  the Be abundances 
 for the stars HD 166983 and HD 219617  for which spectroscopic and
photometric gravities differ considerably. 
For most  stars  in our sample the choice of 
the different metallicities and gravities given in Table 2 move
the points along a constant [Be/Fe] which does not make   a big
difference, with the 
 notable   exceptions of 
 HD 219617 and  HD 160617,
which are shifted from the ``Be-- weak'' zone towards the 
``Be--normal'' strip.
For instance HD 219617 would  have [Be/Fe]=0.12 instead of -0.9 if
its metallicity were [Fe/H]=-1.63 and log g=4.05 instead of -1.08 
and 3.44 as provided by
Axer et al (1994).
HD 160617 would have [Be/Fe]=-0.27 instead of -0.56 if we take
[Fe/H]=-2.05 and log g=3.08. The high value found by Gilmore et al (1992)
which would give [Be/Fe]=0.21 follows from a particularly high
value for the gravity, namely log g=3.8. These two
stars are potentially  Be weak stars  depending 
on the real stellar parameters.
Considering the errors involved in the Be 
abundance determination, the whole 
dispersion of the points along the Be-Fe relation may be due to
observational errors,  with the notable exception of three stars
which will be discussed later on.

\begfig 6.7 cm
\figure{4}{
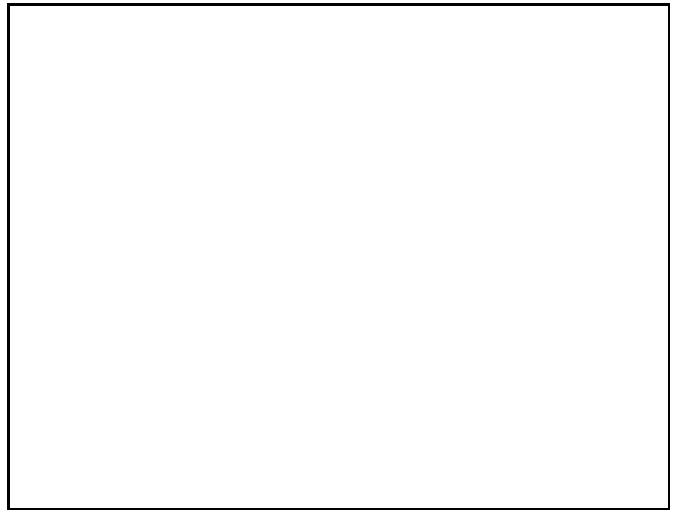
[Be/Fe] versus [Fe/H]. Open octagons are from Table 7 with the exception
of the data from Boesgaard and King (1993) from Table 6. 
Filled octagons are this paper 
}
\endfig

\titlea{Discussion}

 Reeves, Fowler and Hoyle (1970) 
first suggested that spallation reactions between cosmic rays, 
and the interstellar medium are  responsible for  
the production of the light elements Li, Be and B together with their isotopes. 
 A  straightforward  integration over the galactic life of the
present rate of elemental production, i.e. assuming constant
flux and shape of high energy cosmic rays,  accounts for both  
solar Be and B abundances
and  their  ratios, with the remarkable failures of Li abundance
and of isotopic boron   ratio. While for lithium a number of other sources
including the primordial one are viable,  for the B 
isotopes a {\it carrot} of low energy 
cosmic rays was postulated by Meneguzzi et al (1971).

New recent  observations of  abundances for Be, 
$^7$Li, $^6$Li, and B in halo stars have allowed  a   better definition of the 
 evolution
abundance curves for  the light  elements. In the past
Be  was  always  
lower  than in the sun,     
showing a progressive decrease with the 
decreasing of  the stellar  metallicity.
 The  measures
by Rebolo et al (1988), Ryan et al (1991) and Gilmore et al (1991, 1992) 
revealed  a linear increase  of Be with the metallicity.

The precise correlation between Be and Fe contains clues which  
reveal the process responsible for Be
synthesis. 
Spallation reactions may occur when p and $\alpha$ particles collide with 
the CNO atoms at rest in the ISM, or when the fast  
CNO nuclei of the cosmic rays
collide with the p and $\alpha$. The second process produces
fast light elements which may escape from the Galaxy;   this  has been
generally assumed  to produce a modest 
contribution to the synthesis of the light elements.
 
 If the former mechanism is the dominant one 
and the cosmic
rays are produced by supernovae, then  
the Be   abundance is expected to 
be proportional to the square of the metallicity.
To account for the linear behaviour of Be versus Fe, Ryan et al (1990)
and Gilmore et al (1991)
 suggested that Be 
is synthesized in the immediate surroundings of supernovae by spallation
of fast nuclei of C and O ejected by the supernovae against
the protons of the interstellar medium. This idea has been
also further elaborated by Feltzing and Gustafsson (1994) and Tayler (1995).

\titleb{Be-Fe correlation}

The  Be  abundances derived in this paper versus [Fe/H]
 together with all  Be 
observations available 
in literature are shown in Fig. 3.
When stars evolve off the main sequence, 
the surface abundances of Be 
dilute considerably because of the deepening of the surface convection
zones into Be-free layers. 
Standard models predict 
 significant
depletion  on the subgiant 
branch when \t~  $\le$ 5700-5500 ~K (Deliyannis et al 1990, Chaboyer 1993).
This has important implications for the interpretation of Be observations
of some of the stars in our list, namely HD 3795, HD 128279, HD 140283, 
HD 160617, and possibly HD 200654.
 No detectable Be
is found in some of the subgiants,  
 showing that considerable  Be depletion
must have  occurred in these stars. 
 To avoid contamination of the sample abundance 
from evolution effects we have cleaned the  sample from evolved
stars (logg $\le$ 3.6)  unless their  \t~  is $>$ 5750 K.
As a precaution  we kept only the stars for which  all  
parameters from different authors or derived by the
different means fulfill the requirements of gravity and metallicity.
 The Be data show a considerable scatter  
for [Fe/H]$>$  -0.8, which  is 
probably  due to stellar depletion, 
thus for safety we considered only stars with 
lower metallicity. 
 Moreover, we rejected  HD 219617 for which we have a $\Delta$[Fe/H]=0.6 dex,
 HD 189558 
 because of the large (1 dex) 
 difference of  Be abundance between  two 
 different sources, HD 166913 for the large discrepancy in the gravity.
We  also did not consider
 the  upper limits for Be and took the weighted
average of the Be abundances for stars for which multiple measures
are available ( 4 stars: HD 64090 and 
HD 94028, HD 76932, HD 134169).

After this cleaning  we are left  with a sample of 19 stars (see Fig. 5)
in the interval $-2.7<[Fe/H]<-0.8$, for which  we performed a
 $\chi^2$ analysis   taking into account  
 errors both in the Be 
 abundances and in metallicity. 
 The latter are assumed to be $\pm$0.1 dex for all the stars.
The analysis  gives:
\bigskip
 \centerline{ [Be]=1.81($\pm$ 0.114) +1.07($\pm$ 0.08)[Fe/H].}
\bigskip\noindent
with a  reduced $\chi^2$ is 0.72 corresponding to a goodness of fit Q=0.78.
The correlation between Be and Fe is thus   slightly steeper than    the 
[Be] $\propto$0.8[O/H] originally found  by Gilmore et al (1992) from 
only six  determinations, but it is  slightly more gentle than 
the [Be]$\propto$1.256[Fe/H] derived by 
Boesgaard and King (1993).
In general the new Be abundances and  the
repeated analysis of the halo stars ([Fe/H]$<$-1.0) by Boesgaard and
King (1993) support a tight linearity between  Be and Fe.

If we confine the $\chi^2$ fit to the data points with [Fe/H]$<$-1.4 
we obtain:

\bigskip

\centerline{ [Be]=2.89($\pm$0.96) + 1.57($\pm$0.436)[Fe/H]}
\bigskip\noindent

 With
a reduced $\chi^2$ of 0.39 and Q=0.87.
If we confine to [Fe/H]$<$-1.6, which practically means not considering 
one more star, we obtain  [Be]=4.40$\pm$3.91+2.28($\pm$1.26)[Fe/H].
\par
This might suggest that  in the early galaxy a steeper increase
of Be with the metallicity is still possible. 
However, the small number of data  points does
 not allow firm conclusions.  The large error we get in the slope
makes it marginally consistent  with a slope of 1.

  Be spallation involves mainly oxygen, because of its larger  
spallation cross sections with protons, and several authors discussed the
behaviour of Be with 
respect to oxygen (Gilmore et al 1991; Boesgaard and King 1993). 
However,  oxygen abundances are not always 
available for the sample stars, 
and if available  they often do not have  the desired accuracy.
Therefore, instead of using the measured oxygen abundances for the
data sample we investigate the Be versus O 
correlation by using  the O-Fe parametrization
which has been deduced by 
general observations of halo and disk stars, namely we took  
[O/Fe]=0.5 
for   [Fe/H]$<$-1.0,  and [O/Fe]= -0.5[Fe/H] for [Fe/H]$>$-1.0. 
The Be abundances versus oxygen are 
shown in Fig. 6. 
The analysis for stars with 
[O/H]$<$-0.6, which corresponds to [Fe/H]$<$=-0.8,  gives:
\bigskip
\centerline{[Be]=1.38($\pm 0.13$) + 1.13($\pm 0.11$)[O/Fe] }
\bigskip\noindent
 with a  reduced $\chi^2$ is 0.59
and   a goodness of Q=0.76.

 A  fit to the whole sample
gives  a linear relation with a similar slope, slightly offset, but with an 
inferior goodness of  fit ought  to the dispersion in the Be data
at disk and solar metallicities. 
It is worth noticing that if oxygen is increasing
towards lower metallicities,  as  suggested by some observations,
then the correlation 
would result steeper. 
Assuming [O/Fe] increases linearly towards
lower metallicities from [O/Fe]=0.5 at [Fe/H]=-1.0 up to +1.0 at [Fe/H]=-3.0,
the fit now gives a correlation:
\bigskip
\centerline{[Be]=1.13($\pm 0.11$) + 1.38($\pm 0.13$)[O/Fe] }
\bigskip\noindent
with a  reduced $\chi^2$ of  0.39
and  a goodness of  Q=0.94.
 opening the possibility to a faster increase
of Be at lower metallicities.
 In  Fig. 6
the crosses and open squares  show the different behaviours with  [O/Fe]
constant and increasing towards lower metallicities.

 The fit obtained for the stars with  
 [O/H]$<$-0.6 intercepts the upper envelope of [Be] 
 abundances and the meteoritic
value. 
This is also true considering the Be versus [Fe/H], where the extrapolation
at [Fe/H]$\approx$ 0,  is about 
0.2 dex above   the meteoritic value.
If the dispersion observed at solar metallicities is due to
stellar depletion, in opposition to an intrinsic pristine dispersion,
then   the upper envelope of the [Be]
abundances is more representative of the evolution
abundance curve  of Be. 
And if this is the case, from Fig. 6  it is clear 
that  the Be abundance tracks   oxygen closely and  [Be/O] 
remains  always at about the solar   value at any metallicity.
This is  a remarkable output if we consider that that Be production
is related to quantities 
such as  the flux of cosmic rays and  the 
amount of astration   which do not affect oxygen production.
 
Alternative 
nucleosynthesis for 
Be such as that caused by the flood of neutrinos passing through the mantle
and helium core  during explosions of massive stars might also result in a
linear output. 
Malaney (1992) 
suggested that $^9$Be  might be 
produced by $^7$Li(t,n)$^9$Be in the helium shell of 
low-metallicity stars. However,
 no appreciable fraction of the solar abundance of $^9$Be is made
in such a way, and the ejected abundance of $^9$Be 
in metal deficient stars, is quite small if compared with other
isotopes like $^{11}$B.
The process of $\nu$-induced nucleosynthesis
may produce 
interesting amounts for  $^7$Li, $^{10,11}$B, 
but only small amounts of $^9$Be (Woosley and Weaver 1995).

\begfig 6.5 cm
\figure{5}{ 
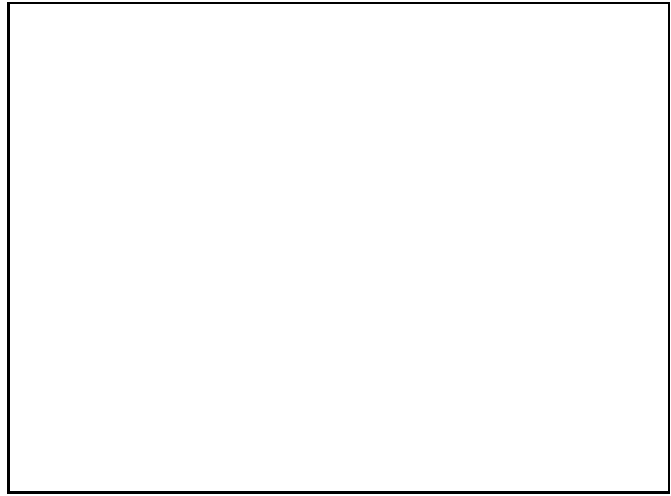
[Be] versus [Fe/H] for a selected sample of stars as described 
 in the text.  
Dashed line is for the  fit of  stars with [Fe/H]$<$-0.9, dotted
line for [Fe/H]$<$-1.4
 }
\endfig

\begfig 6.5 cm
\figure{6}{ 
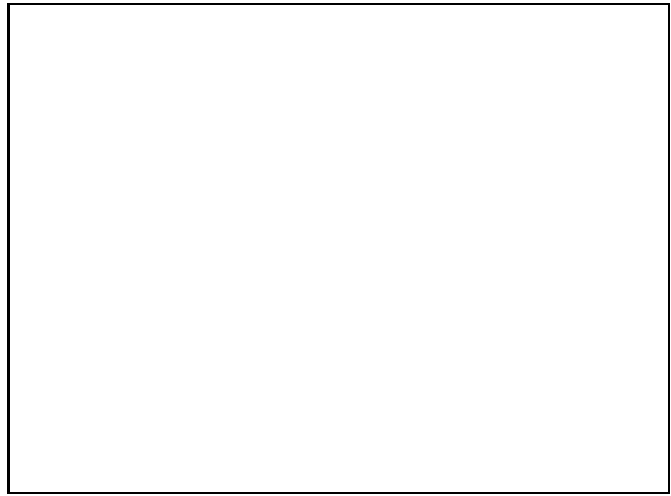
 [Be] versus [O/H]. Crosses are for [O/Fe]=+0.5 in the halo, while
squares are for increasing oxygen towards lower metallicities
as described in the text.  Continuous line is the fit 
for crosses   with [O/H]$<$-0.6 and 
dashed line for all the sample.
  Dotted line is the fit of the squares for [O/H]$<$
 -0.6 and then extrapolated towards solar values
}
\endfig

\titleb{Primordial Be}

The possibility of primordial production of Be in a strongly inhomogeneous
universe was explored by Malaney and Fowler (1988), Boyd and Kajino
(1989) and Thomas et al (1993). Besides the fact that, on theoretical grounds,
large inhomogeneities are unlikely to occur, the presence of a plateau of Be
at low metallicities such as the Spite plateau for Li would be a strong case
for primordial Be production, although not  totally univocal as argued by
Yoshii, Mathews and Kajino (1995).
In Fig. 5 in correspondence of the most metal-poor
objects, there is no evidence for a {\it plateau}  in the Be abundances
analogous to that observed for Li abundances, at least down to [Fe/H]=-2.5
and [Be]=-1.0.
Boesgaard (1995) argued that a plateau   at
[Be] about -1.0
is plausible
but far from conclusive for the lowest metallicity stars HD 140283 and
BD+3 740. 

HD 140283 plays  a special role in the 
existence of a plateau at low metallicities. 
 Our determination
differ  from the previous ones mainly for the higher Teff and the inclusion
of $\alpha$ enhanced opacities in the atmospheric model. As already discussed, 
a further
difference with respect to the Boesgaard and King (1993) analysis
concerns overshooting.
 If the star is a relatively cool subgiant with
 Teff=5660 K, as taken 
 by Boesgaard and King (1993) and Gilmore et al (1992), standard stellar
evolution models predict  $^9$Be  depleted by $\approx$ 0.3 dex
(Chaboyer 1994). Thus
the values [Be]=-0.97$\pm$0.25 (Gilmore et al 1992) and  -0.78$\pm0.14$,
(Boesgaard and King 1993) should be increased correspondingly, producing 
a flattening of Be-Fe relation at the lowest edge.
On the other hand for the \t ~  used
here no post-main sequence depletion is expected and the measured
abundance [Be]=-0.86 should reflect the protostellar one.    It is the 
precise value of \t ~ which determines 
whether the evolution correction has to be applied or not. 

\begfig 6.5 cm
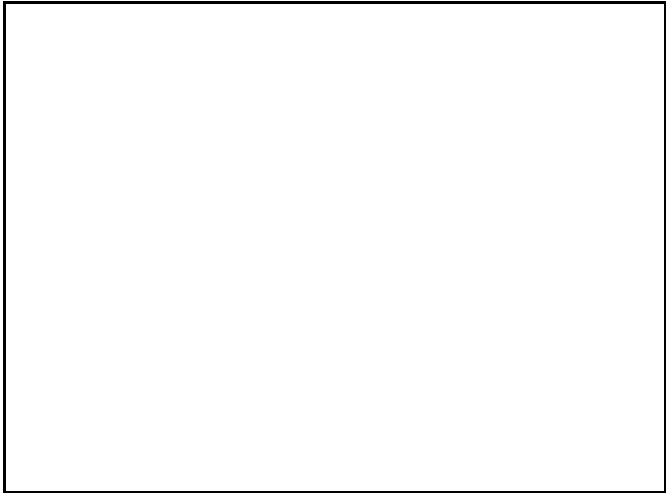
\figure{7}{Residual spectrum 
after  division of HD 76932 with HD 106516 spectra}
\endfig
 A remarkable prediction of inhomogeneous
BBN is a large Be/B ratio of the order of about $\approx$ 100.
So far B has been measured in only three stars
(HD 19445,
HD 140283 and HD 201891) 
with   boron showing  a linear behaviour with iron as Be does. 
 The B/Be 
 ratios have been used to 
 derive information on the origin of the 
 two elements (Walker et al 1993, Olive et al
1994, Fields et al 1995). Most of the arguments rely on
HD 140283, which has  the more reliable
boron determination.
 The LTE [B] =-12.16$\pm 0.14$ by Duncan, Lambert and Lemke (1992) coupled
with 
the [Be]=-0.97$\pm$0.25 (Gilmore et al 1992) or the [Be] -0.78$\pm0.14$,
 Boesgaard and King 1993)
 results in a very
low B/Be ratio, i.e lower than 6.
The result is  exacerbated 
if a post-main sequence correction of 0.3 dex to the 
Be abundance should be applied to 
HD 140283 when the cooler temperature is adopted, with
 B/Be $\approx$ 3. 
  The upwards revision of the Be abundance may be
balanced by a the revision of B abundance by non LTE effects.
Edvardsson et al (1994) and Kiselman (1994)
 have repeated the analysis of the B
abundance taking into account non-LTE effects, leading to an upwards revision
of $\approx$ 0.5 dex, i.e. 0.34$\pm 0.2$.
Therefore, taking into account both revisions,  the  B/Be ratio becomes 
$\approx$ 9, 
still leading  to a marginal  conflict with
the B/Be ratio of 10-20 predicted by GCR  production.
However, for the \t~  we have adopted here  (5814 K) no 
post main sequence dilution  is expected and with our value 
[Be]=-0.86  the   
 B/Be ratio becomes 17, once 
 non-LTE effects for B are accounted for. This ratio  is
in good agreement with the GCR predictions and does not require additional
sources either for  boron or beryllium.

\titleb{Be-depleted  dwarfs}

Be is destroyed in warm stellar interiors and its presence in stellar
atmospheres has important bearings for the study of the stellar outer
layers (Bodenheimer 1966). 
Standard models predict negligible Be depletion in dwarfs 
with \t~  $\ge$ 4900~K, in which the base of the convection zones
is not hot enough to burn beryllium, and  in subgiants with Teff$>$ 5700-5500 
(Deliyannis et al 1990, Deliyannis and Pinsonneault 1990).

In fact, observations by
Garcia Lopez et al (1995) in Pop I Hyades dwarfs show
a plateau in the Be abundances down to  Teff $\approx$ 5200 K, with a decline
afterwards. In these   
stars Li is partially 
depleted  implying that the mixing mechanism 
operating below the convection zone must be efficient enough in transporting 
material down to the Li 
burning layer (T=2.5$\times$10$^6$ K), but not to the Be
burning layer (T=3.5$\times$10$^6$ K).
\par
Extra mixing due to diffusion and rotation may amplify the elemental depletion
if compared to the standard model. Rotationally induced  mixing leads to
a Be depletion strongly dependent on the \t ~ 
and on the degree of 
initial angular momentum (Deliyannis and Pinsonneault 1990). 
We verified that accounting for such a depletion does not 
 destroy  the
linear Be-Fe relation, but this may be due to 
the small number of hot halo stars
available. It will be interesting to repeat the test
when more such stars are  available.
\par
No detectable Be is found  for HD 106516, HD 211998, HD 3795. 
The corresponding upper limits 
 are far below the mean trend of the Be-Fe relation. 
 HD 106516 is a dwarf  with temperature
5995 K and  no depletion is predicted by standard models
for such a temperature. 
The upper limit in the Be is about 1 dex below the average Be-Fe
 relation.
  HD 106516 is a star with properties
very similar to HD 76932. They have the same metallicity,
gravity and almost 
the same effective 
temperature. HD 106516 may actually be  300 K hotter
than HD 76932 according to the colour temperatures in
column 2 of Table 2, but also slightly more metal--rich,
according to the metallicities of Edvardsson \al ~ (1993)
(column 9 in Table 2). 
Whichever the case the only 
remarkable difference between their spectra
is the lack of Be lines in HD 106516.  
This can be appreciated from  Fig. 7 where the spectrum resulting
from the division of the two stars  is shown.
In the division of    their spectra  
all the spectral features cancel out with the exception of the
BeII lines of HD 76932, which clearly show up in  Fig. 7.
Note that a 300 K increase in the temperature of HD 106516
would result in an increase  of only 0.03 dex in the [Be] abundance.
 HD 106516  shows also no Li 
 (Hobbs \& Duncan 1987)  and therefore  the two elements are
  consistently depleted. If we consider the high temperature
of this star, it is likely that
 rotationally induced mixing or diffusion may be  responsible
for the elemental depletion of both Li and Be. 

 HD 3795 has $\log g$ =3.6-3.9, and Be$<$ -0.5, at 
a metallicity where the Be is typically
$>$ 0.5. Li has been measured at less than 0.60 by
Pasquini et al (1994), i.e more than 1.5 dex below the Spite plateau,
and considering that   Teff is $\approx$ 5400 K , convection
 might have  depleted Li according to the standard theory.
However, we do not expect that   Be  has 
been depleted at this temperature.  
Therefore, other mechanisms should be at work  in depleting Be  
from the atmosphere of this star.

 HD 211998 has Teff = 5338 K, no Be  and Li measured 
with [Li]=1.1 (Maurice \al 1984), 
i.e. $\approx$ 1 dex below the plateau value.  
According to Deliyannis and Pinsonneault (1990), 
since Li burns at lower temperatures, 
any significant destruction of Be  
would imply Li to have been completely destroyed.
Thus in this case both the absence 
of Be and  the  behaviour of Be and Li
are at odds with the predictions of standard stellar evolution theory.
However, 
 HD 211998 might be a post turn off star since 
logg=3.26$\pm$ 0.12 from Axer et al (1994), and 3.5$\pm$ 0.26 
in the photometrical
derived gravity as reported in the fourth column of Table 2.  The presence
of Li would be consistent with
HD 211998 being an 
evolved star, since many 
subgiants show lithium at this level
(Pilachowski et al 1993;  Pasquini and Molaro 1996). 
Be dilution is also expected
to be about 0.8 dex and therefore it will be very interesting 
to see whether the real   
Be abundance  is very  much below the present upper limit.

\par
The cool (Teff $\approx$ 5000 K)
halo dwarf Groombridge 1830 (HD 103095) has  a Be abundance  by at least a
 factor 2-3 below   the mean trend and 
 Li has been detected in this star with Li=0.27$\pm$0.06 
 (Boesgaard and King 1993; Deliyannis et al 1994). 
Various possibilities such as an intrinsically low 
beryllium abundance, mass loss or slow mixing have been discussed by 
Deliyannis et al (1994).

The three new stars  given here together with 
Groombridge 1830 are the first cool Be-poor stars,
and measurements of light elements, including boron, are potentially 
very important 
because they can help to discriminate among the various
mechanisms for light element depletion  proposed for cool stars so far.

\titlea{The Li/Be ratio}

The comparison between observed abundances and their theoretical ratios is 
a good test of models of elemental nucleosynthesis.
Since Be has probably a univocal source in the GCR,
observations of Be  help to
constrain the degree to which $^7$Li, $^6$Li
$^{11}$B and $^{10}$B  may have been
produced in cosmic ray
collisions rather than in stellar sources, neutrino nucleosynthesis,
$\alpha$ -$\alpha$ reactions, and  in the Big Bang. 
 The amount of Li produced by spallation of high energy cosmic rays
can be inferred from the observed Be by taking the 
theoretical ($^6$Li+$^7$Li)/$^9$Be 
 derived by Steigman and Walker (1992). The presence of $^6$Li has to be
considered because, observationally, it cannot be split from $^7$Li.
However, $^6$Li is rather fragile 
and is not expected to survive in halo stars cooler than $\approx$ 6300 K
(Brown and Schramm 1988). Since
 $^6$Li has been detected in HD 84937, which has \t $\approx$ 6200 K
(Smith et al 1993),  we take the ($^6$Li+$^7$Li)/$^9$Be =13.1  
only for stars hotter than 6200 K and  $^7$Li/$^9$Be=7.6 for
 the cooler stars, which   should have 
 had their original $^6$Li depleted and therefore 
not contributing to the Li EW at 670.7 nm.
These ratios hold strictly for a 
relative chemical composition of the 
solar type and for the present day spectral distribution
of the cosmic rays. However, changing the chemical composition to account
for the enhancement  of O relative to C and N in the past would not
significantly affect the
results shown here.

\begtabemptywid 19 cm
\tbl{
 Be abundances for halo stars:
 1) this paper, 2) Garcia-Lopez et al (1995) 
3) Boesgaard \&  King (1993) 
 4) Boesgaard \& King revised 5) Ryan et al (1991), (1992) 
6) Gilmore et al (1992),
Rebolo et al (1988). Li abundances  are from  Boesgaard 
\&  King (1993),  Molaro (1991), or from the literature as specified in
the text}
\endtab

The Li abundance for the stars for which Be has been measured are given in 
Table 7.  The 
amount of Li$_{GCR}$  is negligible for low metallicity, where the primordial
fraction and  $\alpha$-$\alpha$ fusion reactions are the  overwhelming 
sources for $^7$Li. Li$_{GCR}$  is about 1\%
at [Fe/H]$\approx$=-2.5 
and 10\% at [Fe/H]=-2.0. For [Fe/H]$>$-1.0 it becomes progressively
more important  and 
 comparable to that of other sources. It might be possible that
the GCR contribution to  present day 
Li is in fact greater than 
what  is generally 
assumed (i.e. $\approx$ 10\%). The relative contributions
are   related to the detailed synthesis
of the two elements during the evolution of the Galactic disk, and
the global uncertainty of these predictions 
leave enough room for such a possibility.

It is remakable that at [Fe/H]$>$ -1.0  in several cases  
  Li$_{GCR}$  
exceeds  the Li which is actually observed.
 The most probable 
explanation is that in these stars Li has been 
largely depleted.
Li   burns at lower temperatures than Be and   
therefore these are probably the 
stars in which Li has been depleted, but not Be.
The sun itself should  be  one of these cases, with 
$2.3\le {\rm [Li]}_{GCR}\le 2.52$, 
which is much greater than the  measured value, i.e.
[Li]= 1.16.
Thus,  Be determinations  offer a powerful  way to pick up stars where
strong Li depletions occurred. In fact there are four stars that have 
temperatures  between 4900 K and 5500 K. 
 For 
stars within this temperature range 
standard models predict Li depletion 
by stellar convection, but not for Be, and this shows the reliability of this
approach to get information on Li depletion.  These stars are shown in 
Fig. 8   as crossed squares.
Of course the method  fails at lower metallicities 
where most of the  Li is supplied by other sources, or when
 Be may be also depleted. It is reliable only   to discriminate  stars
with large Li depletions, and these place 
therefore  a lower limit to the number of real cases.
 
It is interesting to note that 
these objects  where  Li$_{GCR}$ $>$ Li$_{obs}$ occupy
a well specified region in  the [Li]-[Fe/H] 
diagramme. They are shown as crossed circles 
in  Fig. 8 and  these are 
the points which contribute to the dispersion in the Li
data at metallicities [Fe/H]$>$-1.0.
The remaining stars are those which 
are not depleted at all or only 
moderately,  and are preferentially located on the upper
envelope of the Li-Fe diagramme. 
This implies  that the dispersion observed 
in the Li abundances is due  to stellar depletion rather than 
 to intrinsically different protostellar 
Li abundances.  
It  also strongly suggests that the envelope of the data points
in Fig. 8 is  actually  tracing  the increase of Li during
the Galactic life from the 
primordial plateau value up to the present
values, with strong 
implications on the primordial origin of Li.
 Depletion mechanisms of Li, such as those of rotational mixing 
or diffusion, which are currently invoked to deplete an original
high primordial Li,  should have been highly tuned not only to
reproduce a flat plateau for [Fe/H]$<$-1.0, but 
also to make a differential
depletion at higher metallicities which may 
reproduce a 
monotonic increase of Li from the halo value to solar metallicities.

\begfig 6.5 cm
\figure{8}{
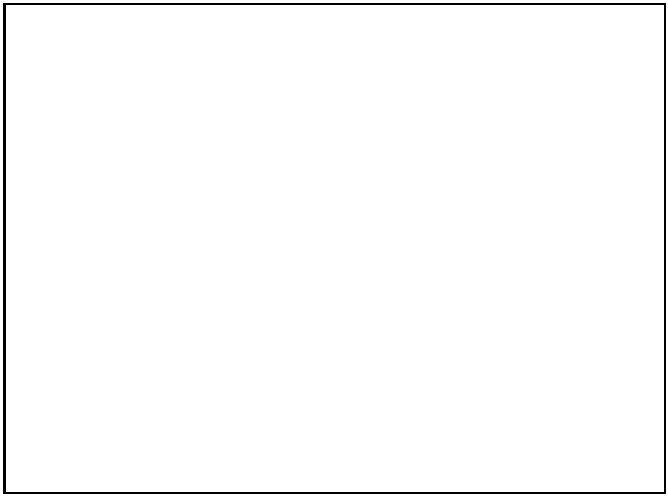
[Li] versus [Fe/H].
Octagons  show stars which are likely Li depleted,
and squares point dwarfs 
 with  5600 $>$ $T_{\rm eff}$ $>$ 
5000 K for which Li is expected depleted but not Be
}
\endfig

\titlea{Conclusions}

Observations at high resolution of 14 metal poor stars are analyzed for
deriving the Be abundance through the 313.1 nm BeII resonance lines.
The analysis has led to the following main conclusions:
\item
{a)} the Be abundance is subject to a small dependence 
of the order of $\approx 0.1$ dex, from the atmospheric model
used due to the particular treatment of
overshooting and/or of the $\alpha$--elements enhancement 
in the chemical 
composition of
the atmosphere;
\item
{b)}
the stellar parameters remain rather uncertain and in particular the surface
gravity which is responsible 
for the main component of the error associated
with the Be abundance;
\item
{c)}
after retaining the best Be determinations in the literature and  after
homogenization to the same set of atmospheric models, Be shows a tight
correlation with Fe with a slope of 1, although a steeper correlation cannot
be ruled out at lower metallicities;
\item
{d)}
 after replacing Fe with O, by using the parametrization which comes
from the general study of the metal poor stars, Be shows a strong
correlation with
oxygen which holds up to  solar values, without showing  evidence
of any break at the formation of the Galactic disk;
\item
{e)}
no evidence of any primordial Be plateau  is found at least down to [Fe/H]=-3.0
\item
{f)}
 few stars are significantly Be deficient, and if they are confirmed to be
truly dwarfs, the absence of Be needs to be explained;
\item
{g)} it is shown in which way
  combined Li and Be observations can be used to pick up
stars which suffered strongly Li depletion. 
These stars occupy 
a specific region in the   
Li versus [Fe/H] diagramme, supporting the interpretation of the
diagramme in terms of a
galactic enrichment of Li from a low Li primordial value of [Li]$\approx$ 2.2;

\acknow{
It is a pleasure to thank A. Gilliotte from La Silla for helping to obtain
these observations, and M. Bessel, F. Spite and F. Primas
for valuable contributions along the course of this work}

\bigskip

\begref{References}
\ref\a Abia C., Isern, J.,  Canal R.:1995 298 465
\ref\gca Anders, E., Grevesse, N.:1989 53 197
\ref\a Axer, M., Fuhrmann, K., Gehren, T.:1994 291 895
\ref\apj Chaboyer, B.:1994 432 L47
\ref\a Chmielewski, Y., Brault, J.W., Mueller, E.A.:1975 42 37
\ref\apj Bessel, M.S., Sutherland, R.S., Ruan, K.:1991 383 L71
\ref\apj Bodenheimer P.:1966 144 103
\ref\apj Boesgaard A.:1976 210 466
\ref Boesgaard, A. 1995 in the proc. of the ESO/EIPC Workshop on {\it
The Light Element Abundances}, P. Crane ed. 363
\ref\aj  Boesgaard A.,  King J. R.:1993 106 2309
\ref\apj Boyd, R.N., Kajino, T.:1989 336 L55
\ref\apj Brown, L. Schramm, D. N.:1988 329 L103
\ref Castelli F., 1996, Proccedings of the Vienna workshop
on Model Atmospheres and Spectrum Synthesis, July 1995, ASP Conference
Series, in press  
\ref\a Castelli, F., Gratton, R., Kurucz, R.L.:1996 in press ~
\ref\aj Crawford, D.L.:1975 80 955
\ref\pasp Crawford, D.L., Mandwewala, N.:1976 88 917
\ref\a  D'Antona, F., Mazzitelli, I.:1984 138 431
\ref\apjsupl Deliyannis, C.P., Demarque, P., Kawaler, S.D.:1990 73 21
\ref\apj Deliyannis, C.P., Pinsonneault M. H.:1990 365 L67
\ref\apj Deliyannis, C.P., Ryan, S.G., Beers, T.C., Thorburn, J.A.:1994 425 L21
\ref\apj Duncan, D.K., Lambert, D.L., Lemke, M.:1992 401 584
\ref\a Edvardsson, B., Andersen, J., Gustafsson., B., Lambert, D.L.,
Nissen, P.E.,  Tomkin, J.:1993 275 101
\ref\a Edvardsson, B., Gustafsson, B., Johansson, S. G., Kiselman, D.,
	Lambert, D. L., Nissen, P. E., Gilmore, G.:1994 290 176
\ref\aj Eggen, O.J.:1987 93 393
\ref\apj Feltzing, S., Gustafsson, B.:1994 423 68
\ref\apj Fields, B. D., Olive, K., Schramm D. N.:1995 439 854
\ref Freytag B., 1996, Proccedings of the Vienna workshop
on Model Atmospheres and Spectrum Synthesis, July 1995, ASP Conference
Series, in press  
\ref\a Fuhrmann, K., Axer, M., Gehren, T.:1993 271 451
\ref\a Fuhrmann, K., Axer, M., Gehren, T.:1994 285 585
\ref\a Garcia Lopez, R. J., Rebolo, R., Perez De Taoro, M. R.:1995 302 184
\ref\a Garcia Lopez, R.J., Severino, G., Gomez, M.T.:1995 297 787
\ref\apj Gilmore, G., Edvardsson, B., Nissen, P.E.:1991 378 17 
\ref\nat Gilmore G., Gustafasson B., Edvardsson B., Nissen P.E.:1992 357 379
\ref Grevesse, N., Noels 1993 in {\it Origin and Evolution of the Elements}
edited by N. Prantzos, E. Vangioni-Flam, and M.
Cass\`e, 143 
\ref\a Hannaford, P., Lowe, R.M., Grevesse, N., Noels, A.:1992 259 301
\ref\asupl Hauck B.,  Mermilliod M.:1990 86 107 
\ref\apj Hobbs, L. M., Duncan, D. K.:1987 317 796
\ref\apj Jedamzik, K., Fuller, G.M., Mathews, G.J.:1994 423 50
\ref\apj Kajino, T., Boyd, R. N.:1990 359 267
\ref\a Kiselman, D.:1994 286 169
\ref Kiselman,D., Carlsson, M.,
1995 in the proc. of the ESO/EIPC Workshop on {\it
The Light Element Abundances}, P. Crane ed. 372
\ref Kurucz R.L. 1993a CD--ROM 13
\ref Kurucz R.L. 1993b CD--ROM 18
\ref\apj Kurki-Suonio, H., Matzner, R.A., Olive, K.A., Schramm, 
D.N.:1990 353 406
\ref\apj Laird J.B.:1985 289 556
\ref\apj Luck, R.E., Bond, H.E.:1985 292 559
\ref\a Magain, P.:1987 181 323
\ref\apj Malaney, R.A., Fowler, W.A.:1989 345 L5
\ref\apj Mathews, G. J., Meyer, B. S., Alcock, C. R., Fuller, G. M.:1990 358 36
\ref\a Maurice E., Spite F., Spite M.:1984 132 278
\ref\a Meneguzzi, M., Audouze, J., Reeves, H.:1971 15 337
\ref Molaro, P.:1987 PHD Thesis SISSA Trieste
\ref\msait Molaro, P.:1991 62 17 
\ref\a Molaro P., Beckman J.E.:1984 139 394
\ref Molaro, P.,  Beckman J. E., Castelli F.:1984 ESA SP-219, 197
\ref Molaro P., Bonifacio P., Castelli F. Pasquini L., Primas F.
1995a in the proc. of the ESO/EIPC Workshop on {\it
The Light Element Abundances}, P. Crane ed. 415
\ref\a Molaro, P., Primas, F., Bonifacio, P.:1995b 295 L47
\ref\msait Molaro, P., Bonifacio, P., Primas, F.:1995c 66 323
\ref\a Nissen P. E., Gustafsson B., Edvardsson B., Gilmore G.:1994 285 440
\ref\apj Olive, K.A., Prantzos, N., Scully, S.,
Vangioni-Flam, E.:1994 424 666
\ref\a Olsen E.H.:1988 189 173
\ref Pagel, B. E. J., 1994 in Lynden-Bell D., ed., NATO advanced research
workshop on Cosmical Magnetism Kluwer, Dordrecht, p. 113
\ref\a Pasquini, L., Liu, Q., Pallavicini, R.:1994 287 191
\ref\a Pasquini, L., Molaro, P.:1996 307 761
\ref\apj Pilachowski, C.A., Sneden, C., Booth, J.:1993 407 699
\ref\apj Prantzos N., Cass\`e, M., Vangioni-Flam E.:1993 403 630
\ref Primas F.:1995  {Tesi di Dottorato -- Universit\`a di Trieste}
\ref\a Rebolo R., Molaro, P., Abia, C., Beckman, J.E.:1988 193 193
\ref\apj Reeves, H., Meyer, J.P.:1978 226 613
\ref\nat Reeves H., Fowler, W.A., Hoyle, F.:1970 226 727
\ref\aj Ryan, S.G., Norris, J.E.:1991 101 1835
\ref\aj Ryan, S. G., Norris, J. E., Bessell, M. S.:1991 102 303
\ref\apj Ryan S. G., Norris, J. E., 
Bessel, M. S.,  Deliyannis, C. P.:1992 388 184
\ref\asupl Schuster W.J., Nissen P.E.:1988 73 225
\ref\a Schuster W.J., Nissen P.E.:1989 221 65
\ref\apj Smith, V.V., Lambert, D. L., Nissen, P.E.:1993 408 262
\ref\a Spite, F., Spite, M.:1982 163 140
\ref\apj Steigman, G., Walker, T.P.:1992 385 L13
\ref\mn Tayler R., J.:1995 273 215
\ref\apj Terasawa, N., Sato, K.:1990 362 L47
\ref\apj Thomas, D.,{Schramm}, D.N., {Olive}, K.A., Mathews, G.J.
    {Meyer}, B.S.,{Fields}, B.D.:1994 430 291
\ref\apj Vangioni-Flam E., Casse M., Audouze J., Oberto Y.:1990 364 568
\ref\apj Walker, T.P., Viola, V.E., Mathews, G. J.:1985 299 745
\ref\apj Walker, T.P., Steigman G., 
Schramm D. N., Olive K. A., and Fields, B.:1993 413 562
\ref\anrev Wheeler, J. C., Sneden, C., Truran, J.W.:1989 27 291
\ref\apjsupl Woosley, S.E., Weaver, T.A.:1995 101 181
\ref\apj Yoshii, Y., Mathews, G.J., Kajino, T.:1995 447 184
\endref
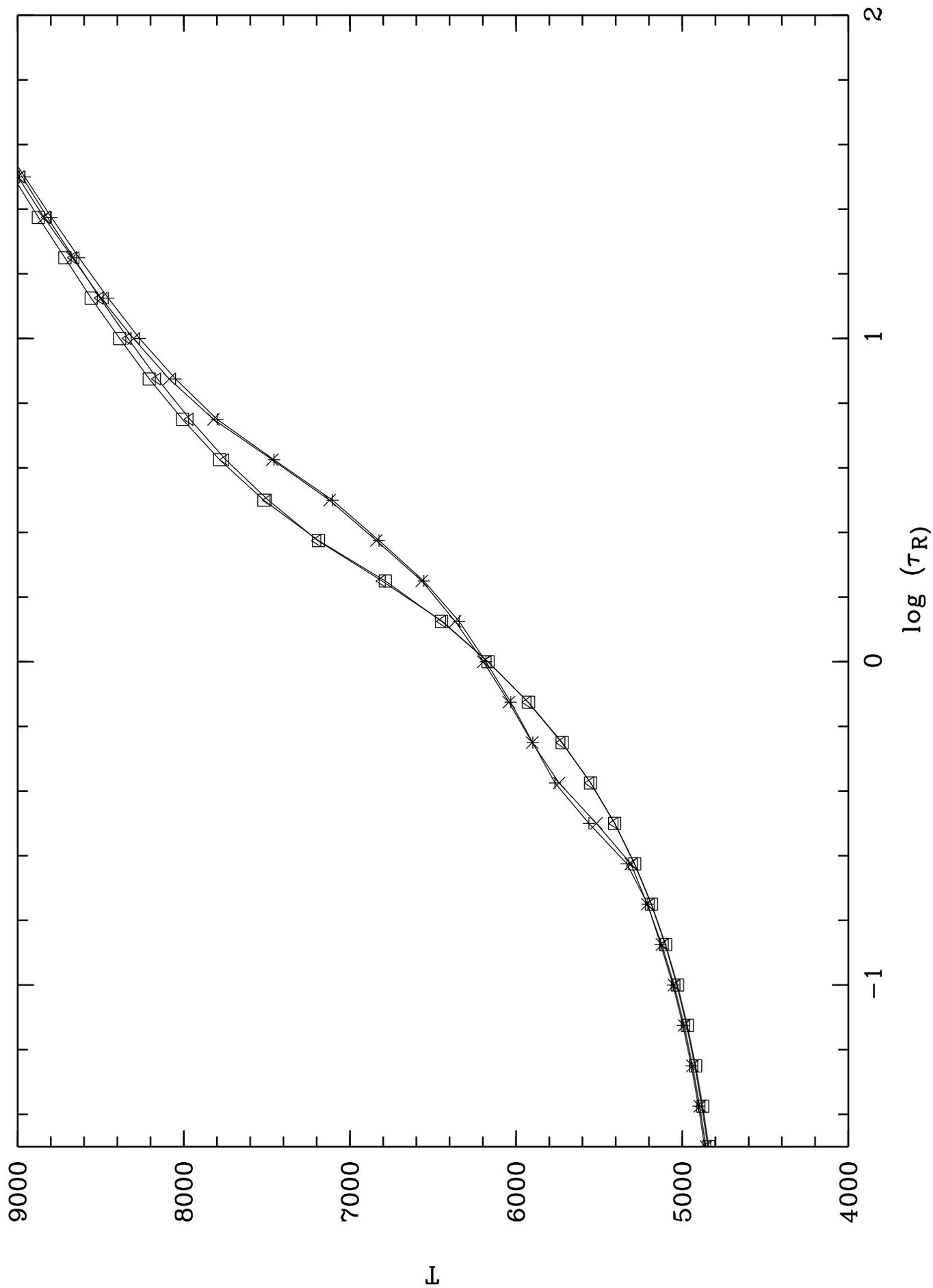
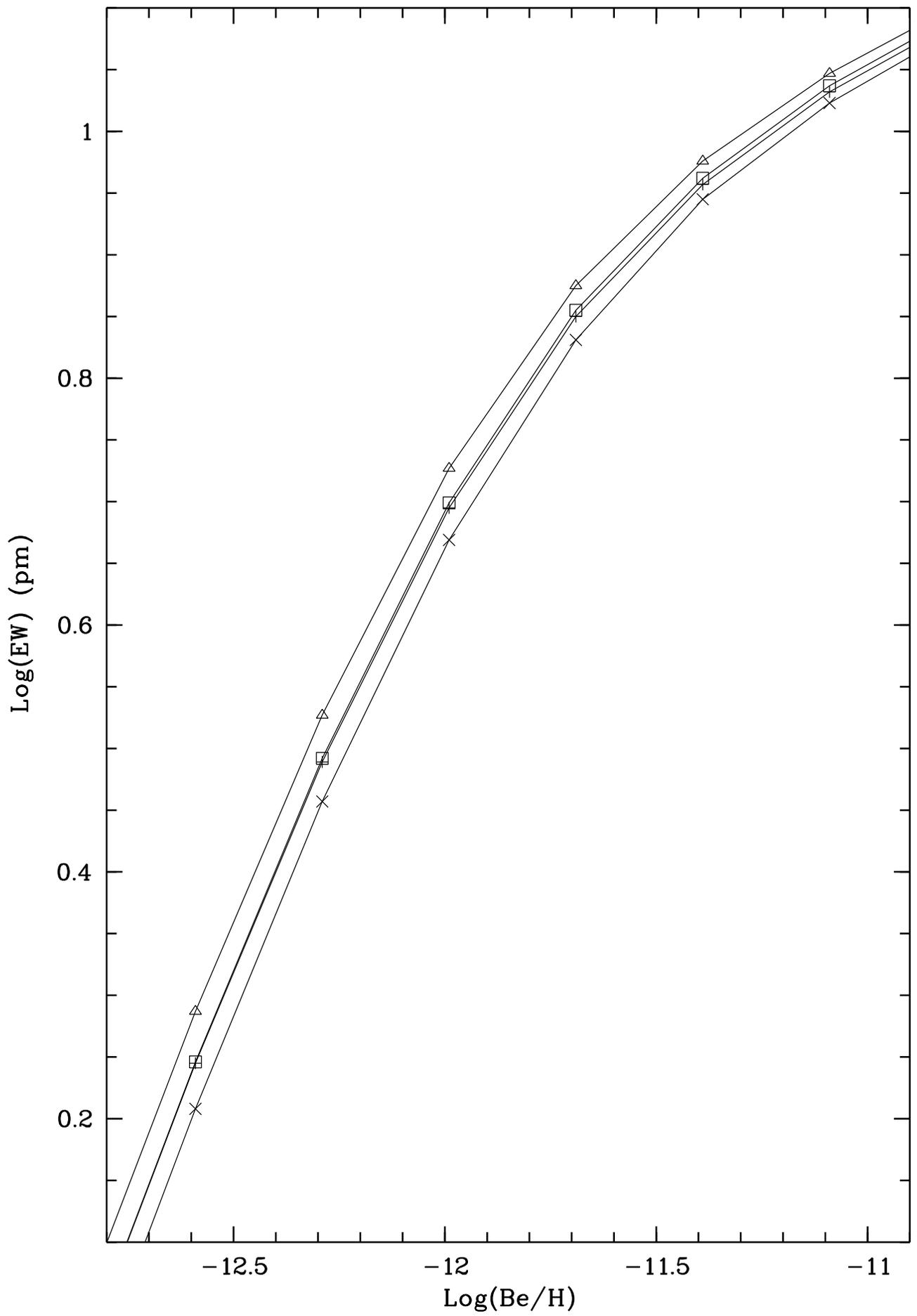
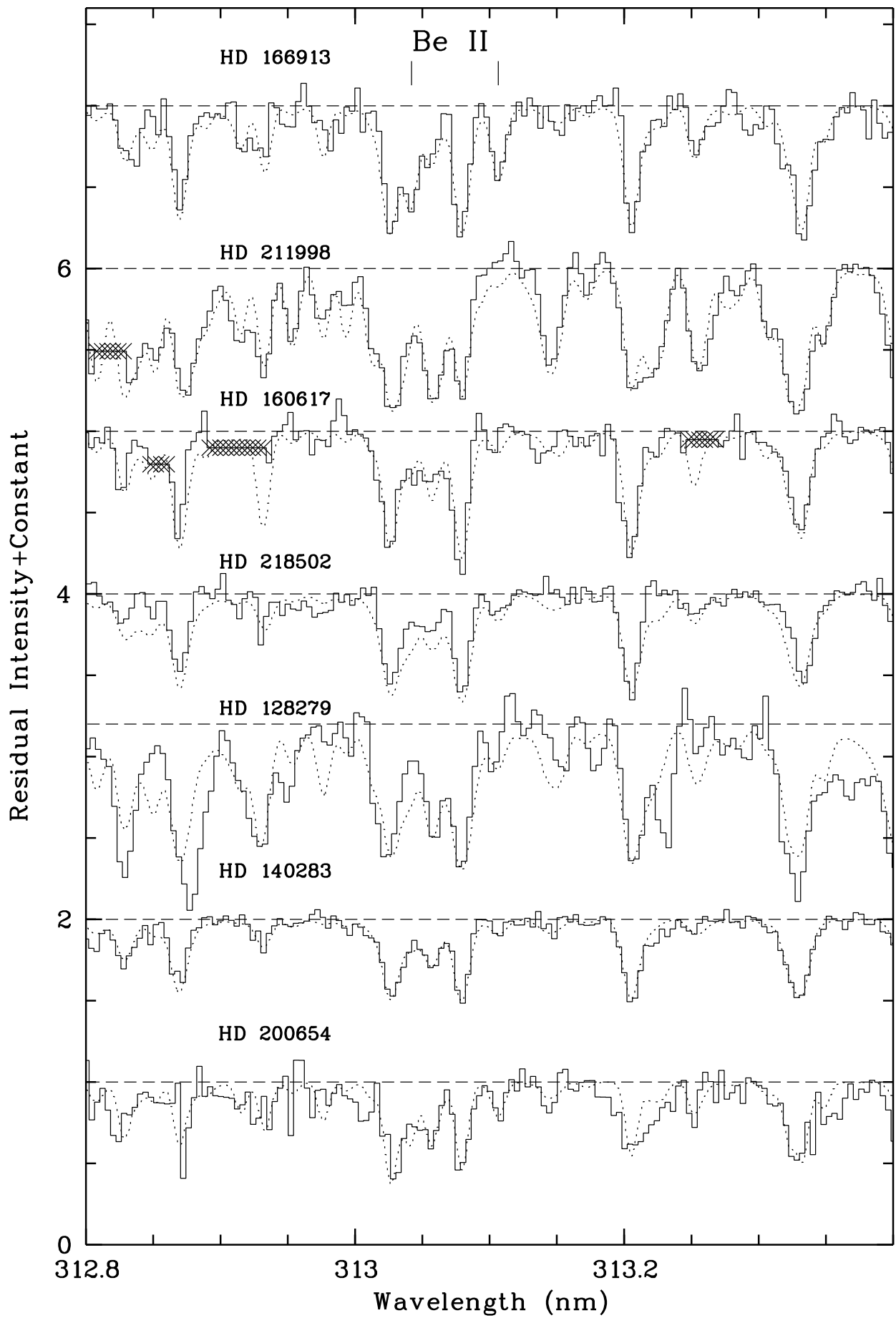
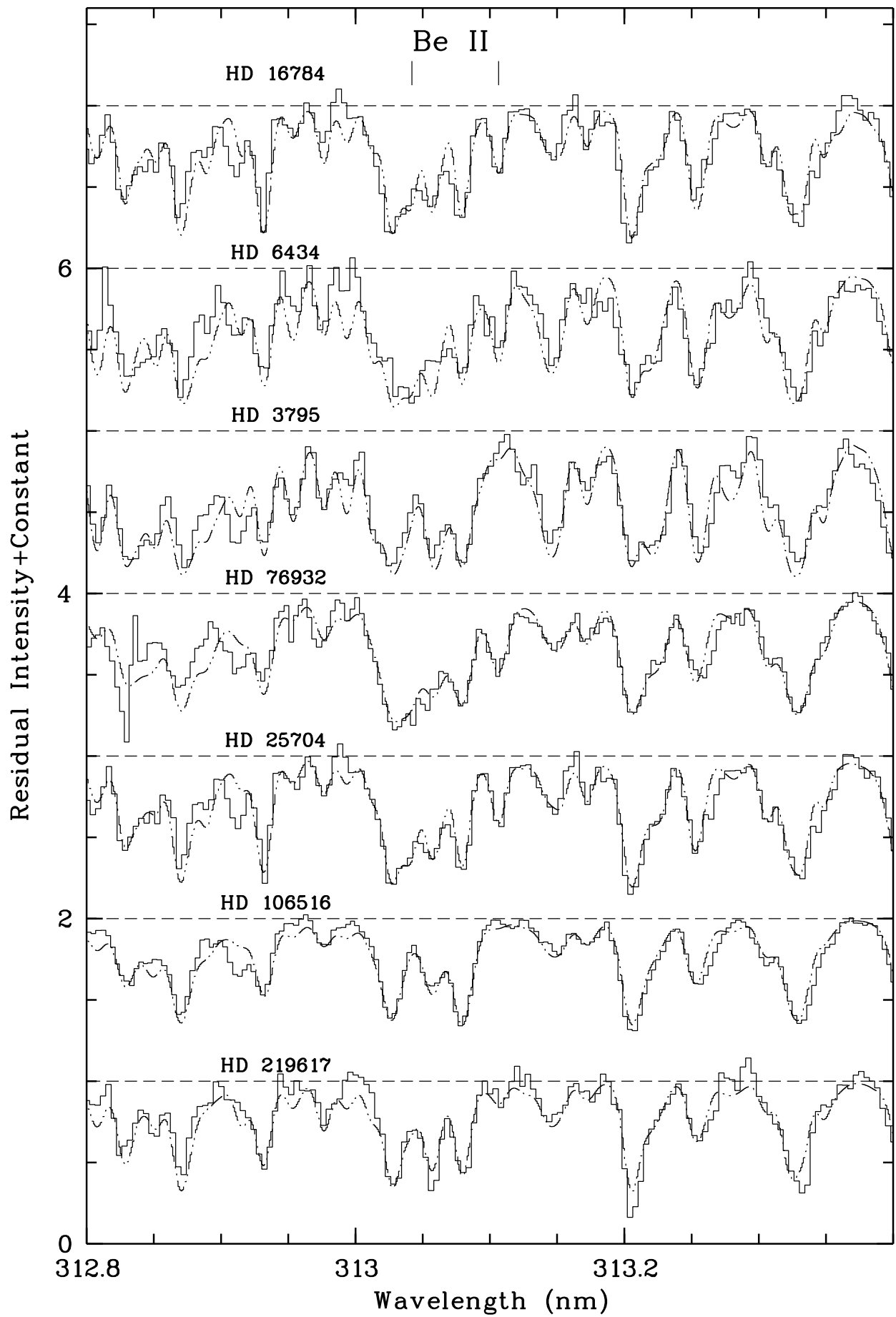
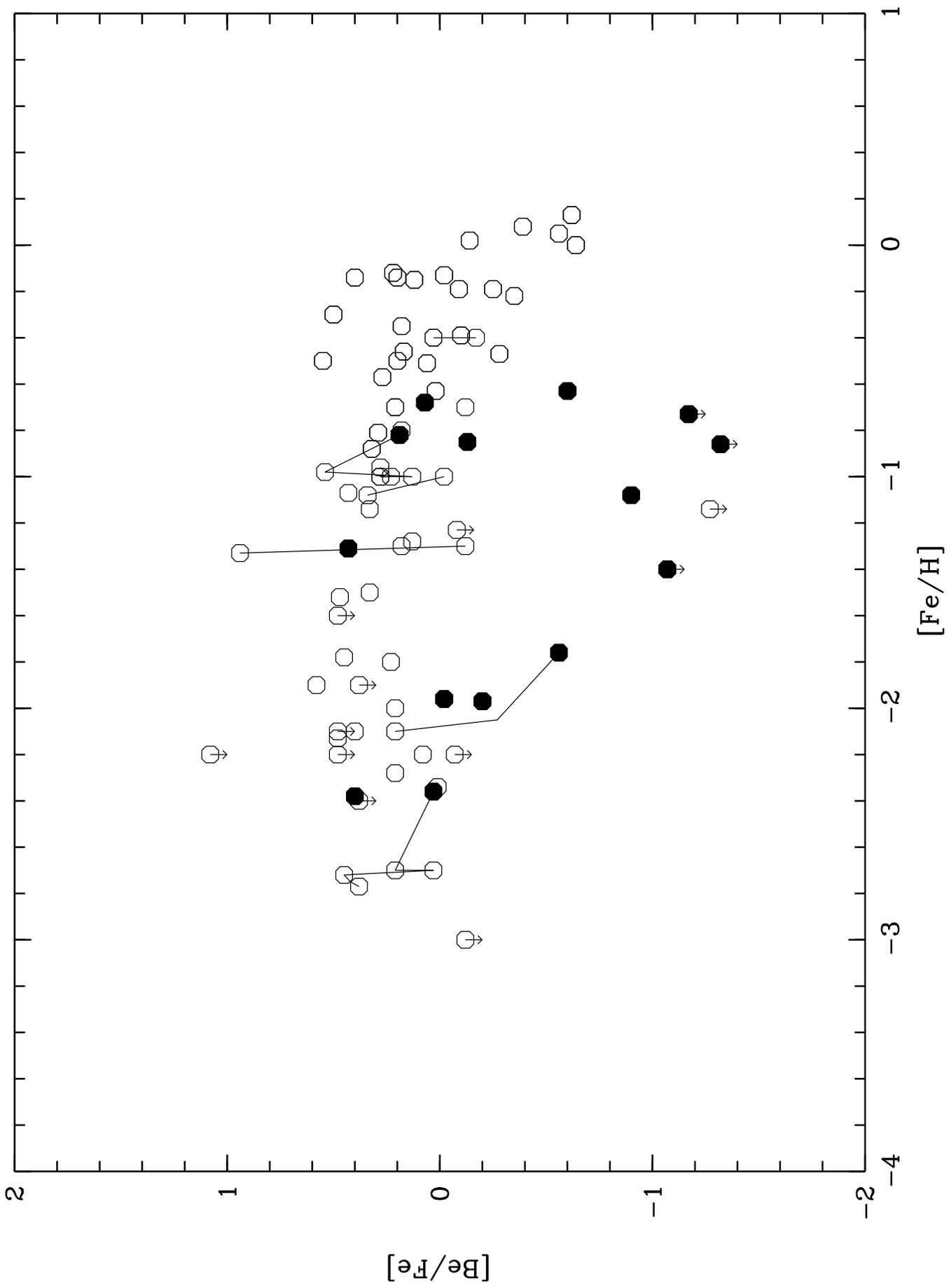
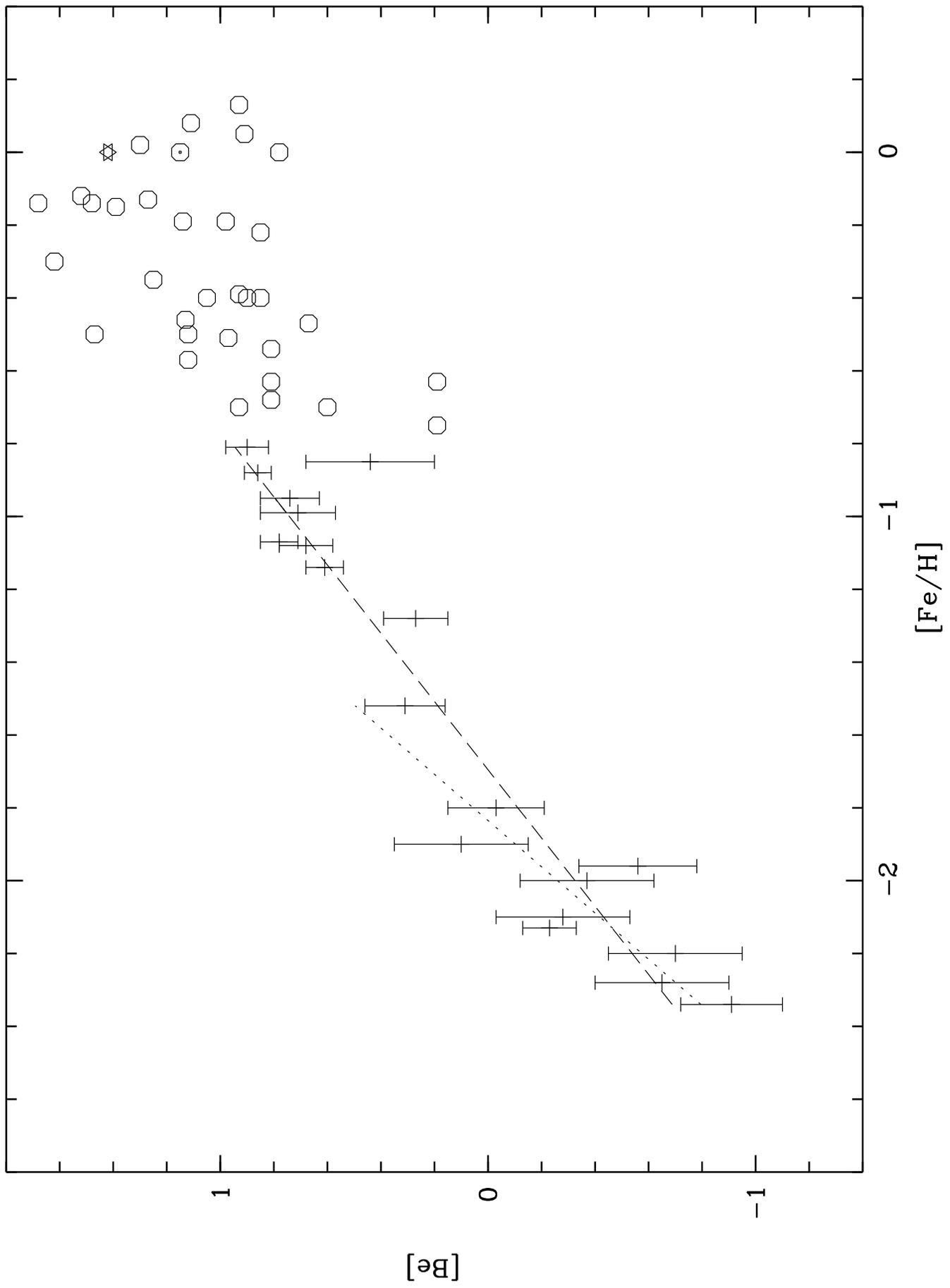
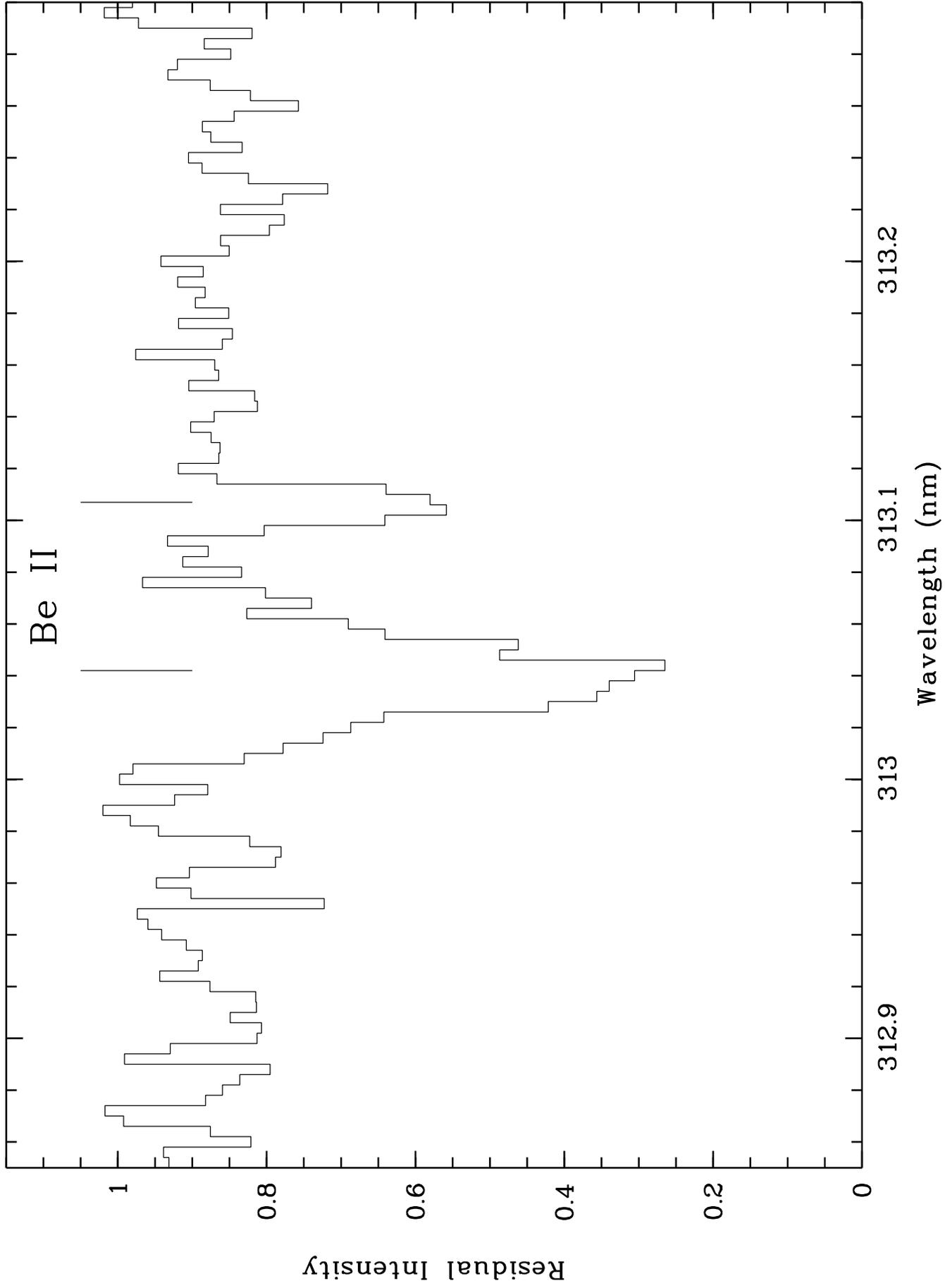
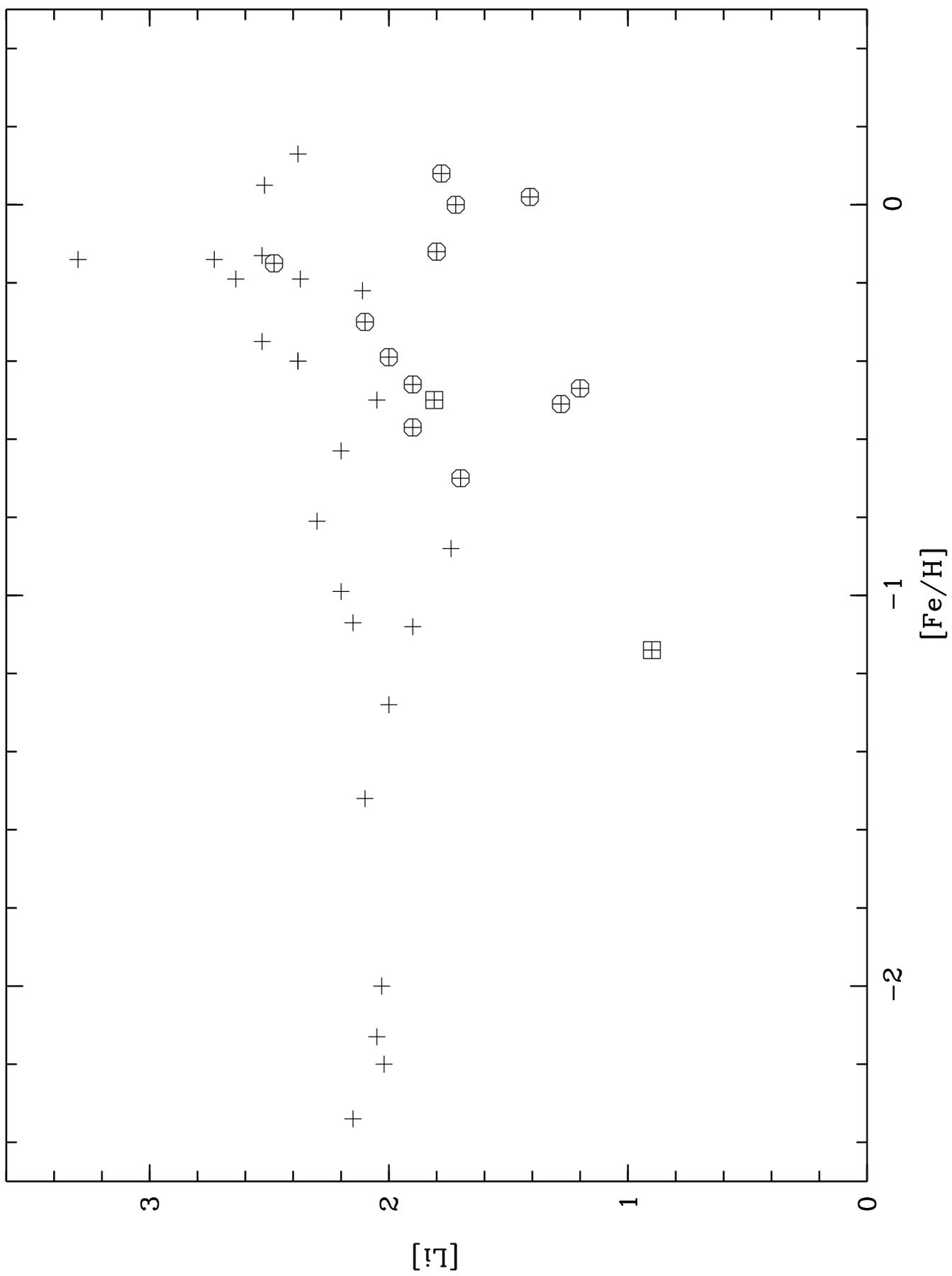
\bye